\documentclass[11pt]{article}
\pdfoutput=1

\usepackage[dvipsnames]{xcolor}
\usepackage{graphicx}
\usepackage{subcaption}
\usepackage{epsfig}
\usepackage{epstopdf}
\usepackage{enumitem}
\usepackage[T1]{fontenc}

\DeclareSymbolFont{myletters}{OML}{ztmcm}{m}{it}
\DeclareMathSymbol{\uplambda}{\mathord}{myletters}{"15}

\usepackage{amsfonts}

\usepackage{hyperref}
\hypersetup{
    colorlinks=true,
    linkcolor=blue!70,
    citecolor=RedViolet,
    filecolor=magenta,      
    urlcolor=cyan,
}

\usepackage{color}
\usepackage{float}
\usepackage{multirow}
\usepackage[toc,page]{appendix} 

\usepackage[mathscr]{euscript}
\usepackage{lmodern} 
\usepackage{amsmath}                                                    
\usepackage{amsthm}                                                     
\usepackage{amssymb}
\usepackage{amsfonts}
\usepackage{mathptmx}
\usepackage{slashed}
\usepackage{latexsym}

\numberwithin{equation}{section} 

\newcommand{\newc}{\newcommand}
\newc{\be}{\begin{equation}}
\newc{\ee}{\end{equation}}
\newc{\bg}{\begin{gathered}}
\newc{\eg}{\end{gathered}}
\newc{\tref}[1]{Table \ref{#1}}
\newc{\eref}[1]{Equation \eqref{#1}}
\newc{\su}[1]{$SU(#1)$}
\newc{\bm}[1]{\mathbf{#1}}
\newc{\fref}[1]{Figure \ref{#1}}

\newc{\ra}{\rightarrow}
\newc{\lra}{\leftrightarrow}
\newc{\ov}{\overline}
\newc{\ba}{\begin{eqnarray}}
\newc{\ea}{\end{eqnarray}}
\newc{\mf}{\mathsf}
\newc{\nn}{\nonumber}

\def\be{\begin{equation}}
\def\ee{\end{equation}}
\def\bea{\begin{eqnarray}}
\def\eea{\end{eqnarray}}

\begin{document}

\begin{titlepage}
\thispagestyle{empty}

                \vspace*{0.7cm}

                \begin{center}
                   { {\bf \Large{Hybrid Inflation, Reheating and Dark Radiation in a IIB perturbative moduli stabilization scenario}}}
                        \\[12mm]
                      Waqas Ahmed$^{a}$~\footnote{E-mail: \texttt{waqasmit@hbpu.edu.cn}},  Athanasios Karozas$^{b}$~\footnote{E-mail: \texttt{athanasioskarozas@gmail.com}}, 
                George K. Leontaris$^{b}$~\footnote{E-mail: \texttt{leonta@uoi.gr}},
                        Ilias Tavellaris$^{b}$~\footnote{E-mail: \texttt{i.tavellaris@uoi.gr}}
                \end{center}
                \vspace*{0.50cm}
                        \centerline{$^{a}$ \it
                                Center for Fundamental Physics and School of Mathematics and Physics, Hubei Polytechnic University}
                        \centerline{\it Huangshi 435003, China}
                \vspace*{0.2cm}
                \centerline{$^{b}$ \it
                         Physics Department, University of Ioannina}
                        \centerline{\it 45110, Ioannina,        Greece}
                \vspace*{1.20cm}
                \begin{abstract}
 We study the cosmological implications of an effective field theory model derived within a configuration of D7 brane stacks in the framework of type-IIB string theory. We consider a suitable geometric setup where the  K\"ahler moduli fields are stabilized and the parametric space is constrained so that a de Sitter vacuum is ensured. In addition to the moduli fields we also take into account the usual Higgs and matter fields
 included in the effective field theory. In this background, we implement the standard hybrid inflation scenario with a singlet scalar field acting as the inflaton and the Higgs states serving as waterfall fields. Radiative corrections and soft supersymmetry breaking terms play an essential role in the realization of a successful inflationary scenario consistent with the present cosmological data. Small tensor-to-scalar ratio values are predicted, which can be probed in future planned experiments. Further constraints on the model's parameters are derived from bounds on dark radiation which is measured as a contribution to the effective number of neutrino species $N_{\rm eff}$. In particular, we find an excess of $\Delta N_{\rm eff}\leq{0.95}$ at $2\sigma$ confidence level with natural values of the involved couplings.

                \end{abstract}
        \end{titlepage}

\thispagestyle{empty}
\vfill
\newpage
{
  \hypersetup{linkcolor=black}
\tableofcontents
}

\section{Introduction }

Cosmological inflation is one of the most successful candidate scenarios in explaining the evolution of the Universe and its large-scale structure observed today. Meanwhile,
numerous effective quantum field theory models have been built with the aim of conciliating cosmic inflation with particle physics models describing the low energy observables. 
  A key criterion on the quest for appropriate models would be  
to have an ultra-violet (UV) completion in a quantum theory of gravity, valid up to the Planck scale $M_P$. 
 In this context, 
at present, string theory appears to be the only promising candidate for a consistent quantum theory at such a
high scale which incorporates the Standard Model (SM) and its minimal supersymmetric extension (MSSM). String theory, however, is
formulated in a ten-dimensional spacetime framework and, therefore,  
compactification of the six extra dimensions is required to 
 achieve a four-dimensional effective field theory compatible
 with the observed world. The reduction of the corresponding 
higher dimensional string action to four spacetime dimensions,
however,  entails an immense set of string vacua which is commonly known as the string landscape.    Yet,
starting from a successful effective field theory that adequately describes the known physics phenomena, we cannot always embed it in a string theory framework~\footnote{See reviews~\cite{Palti:2019pca, Agmon:2022thq, VanRiet:2023pnx}  and references therein.}. 

On the other hand,  consistent effective field theory models emerging after
 compactification should possess a number of important features.   Among them,  they should predict 
  a positive tiny cosmological constant $\Lambda $ of the order $\Lambda\approx 10^{-122}\, M_{P}^4$
    which could account for the dark energy, as suggested by cosmological observations. A simple way to realize such a scenario is within an effective model involving a scalar field $\phi$  with a potential $V(\phi)$ which displays a (possibly metastable) positive minimum equal to the cosmological constant  $\Lambda$.  
  In fact,   
    effective field theory models from strings compactified on Calabi-Yau (CY) manifolds contain a vast number of moduli fields and some of them could play the role of the inflaton $\phi$.  In this context, it is inferred that the cosmological issues are intertwined with the well-known problem of moduli
    stabilization.  In fact,
 moduli stabilization and  (metastable) de Sitter vacua,   play a key role in the 
 successful implementation of the cosmological inflationary scenario in effective field theory (EFT)  models of string origin. Therefore,
 reconciling these two issues is essential in the quest for a suitable non-vanishing effective potential of
 some scalar field enacting as the inflaton $\phi$ which rolls down to the minimum of its (relatively shallow) potential. This enables the required exponential growth of the Universe, provided that the trajectory length  $\sim\Delta\phi$ of the field $\phi$   to reach the minimum is sufficiently long to trigger inflation.

  A particular class of models involves constructions with
  large volume compactification scenarios~\cite{Balasubramanian:2005zx, Berg:2007wt, Reece:2015qbf, Cicoli:2017shd, Cicoli:2021dhg, Gao:2022fdi} and  inflatons associated with the K\"ahler moduli fields  $T_k=\tau_k+ia_k$. 
 In  previous scenarios~\cite{Antoniadis:2018hqy, Antoniadis:2019rkh, Basiouris:2020jgp, Basiouris:2021sdf} considered in the context of type-IIB theory, it was
 shown that the internal volume modulus ${\cal V}$ expressed in terms of the real components of K\"ahler moduli, Re$T_k=\tau_k$,  is a suitable  candidate for this role, ($\phi\propto \log{\cal V}$).  Radiative corrections associated with intersecting space-filling D7  branes on the other hand, provide a stabilization mechanism for  K\"ahler moduli, and an uplift to their scalar potential through their universal abelian factors, thus way leading to a positive cosmological constant~\footnote{There are many other approaches on these issues and a vast literature exists. For an updated list, see e.g. the recent review articles~\cite{Cicoli:2023opf, Leontaris:2023obe}.}. In particular, K\"ahler moduli stabilization is achieved through a non-zero potential generated by $\alpha'$ and radiative (logarithmic) corrections induced when closed string loops
 traverse their codimension-two bulk towards localized gravity sources. Furthermore, the dS vacuum is obtained due to the positive D-term contributions, (originally proposed in~\cite{Burgess:2003ic}) coming from the intersecting D7-branes.

 In  the large volume limit, the induced effective potential for the K\"ahler  moduli  acquires a  simple structure
 possessing two local extrema (a minimum and a maximum) and approaches zero for $\phi\to \infty$.  
 The separation distance of its two local extrema is of the order, $\Delta\phi=\phi_{max}-\phi_{min}\propto \log({\cal V}_{max}/{\cal V}_{min})$.
  In the minimal effective model involving only moduli fields, it can be parametrized in terms of a single non-negative parameter while the largest possible separation $\Delta\phi$  occurs at a critical value of this parameter where beyond this point only AdS solutions appear. 
   A non-zero value of the aforementioned parameter exists at which a new inflationary small-field scenario is successfully implemented. 
In this novel scenario, most of the required number $N_0$ of efolds  ($N_0\sim 60 $) are collected in the vicinity of the minimum of the potential and the prediction for the tensor-to-scalar ratio density fluctuations in the early universe is $r\approx 4\times 	10^{-4}$.  Despite these successes of the model, the picture is not complete and a waterfall mechanism has to be realized in order to end inflation and bring the dS vacuum down to a lower value compatible with the cosmological constant. It has been shown~\cite{Antoniadis:2021lhi} that when open string states appearing in the D7 brane intersections are included in the massless spectrum, appropriate magnetic fluxes and specific brane separations can be chosen in a way that, for  ${\cal V}$  less than some critical value, a charged open string scalar becomes tachyonic.  In~\cite{Antoniadis:2021lhi, Antoniadis:ijmpa} it was shown that such a state can represent a waterfall field. In possible generalizations of this scenario, one may include several of such fields, which end inflation and provide a deeper vacuum in accordance with the present value of $\Lambda$.
 
 In the present work, an alternative scenario is proposed where it is taken into account that, in addition to the moduli fields, the EFT model accommodates the ordinary fermion matter and the Higgs fields in appropriate representations on the gauge group stemming from the specific details of the compactification procedure.  Then, a possible and interesting variation of the scenario described previously would be that the Higgs rolls down a potential hill towards a new lower minimum, where its initial condition is defined around the metastable vacuum of the moduli potential. 
 In the present construction, the metastable vacuum is determined by the K\"ahler moduli and the associated compactification volume  ${\cal V}$.
More precisely, we implement the standard hybrid inflation~\cite{hybrid} with a singlet scalar field acting as inflaton and the Higgses operating as waterfall fields. In this scenario, the vacuum energy is determined by the scalar field and waterfall fields.  We consider radiative corrections which are essential in shaping the slope along the inflationary track. Since supersymmetry (SUSY) is broken during inflation, we also include SUSY soft terms as well. The soft terms play an important role in order to achieve spectral index ($n_s$) values consistent with the current experimental bounds. We find small tensor-to-scalar values that can be probed in future designed experiments. We also discuss dark radiation and show that change in the effective degree of neutrinos satisfies the bound  0.95$\%$ confidence level with natural values of the involved couplings.

The paper is organized as follows. In Section \ref{sec2} we present the basic features of the model including the K\"ahler potential and the superpotential. An analysis of the effective potential is given in Section  \ref{sec3} followed by the presentation of the inflation setup along with our inflationary numerical predictions. In Section \ref{sec4} we discuss reheating and dark radiation predictions. Section \ref{sec5} concludes the paper.

\section{Description of the model and its constituents} \label{sec2}
In this work, we consider a  type-IIB string  framework in ten dimensions where six of them are compactified on a Calabi-Yau  threefold
${\cal X}$.  We restrict our attention mainly to the moduli spectrum and use the following notation:   $\phi$ represents the dilaton field,  while  $T_i$, 
 and  $z_a$  denote the K\"ahler and the complex structure (CS) moduli respectively. Furthermore, we introduce the usual axion-dilaton combination 
\be 
\tau= C_0+ i\,e^{-\phi}\equiv {{ C_0}+\frac{i}{ g_s}}~,
\label{axidil}
\ee 
where $g_s$ is the string coupling and $C_0$ a 0-form potential (RR-scalar).  We assume a perturbative  -flux induced- superpotential $ W_0$ of the form proposed in~\cite{Gukov:1999ya}. At the classical level,  $ W_0$ is a holomorphic function which depends on the axion-dilaton modulus $\tau$, and the CS moduli $z_a$~\footnote{We dispense with the use of non-perturbative corrections which would also introduce the K\"ahler moduli through terms of the form ${\cal W}_{NP}\propto e^{-aT_k}$. As explained in the subsequent analysis, $T_k$ can be perturbatively stabilized through one-loop corrected K\"ahler potential.}. 
 $\tau$  and $z_a$  are stabilized in the standard supersymmetric way, by solving   $D_{\tau} W_0=0,\; D_{z_a}{ W}_0=0$, where $D_I=\partial_IW+W \partial_IK$ are the covariant derivatives.

We consider a geometric configuration of three intersecting D7-brane stacks equipped with magnetic fluxes.  Regarding the K\"ahler potential, we will take into account $\alpha'$ corrections as well as the effects of a novel four-dimensional Einstein-Hilbert (EH) term (localized in the internal space) which is generated from higher derivative terms in the ten-dimensional string effective action~\cite{Antoniadis:2019rkh}. This setup induces logarithmic corrections to the scalar potential via loop effects. Taking into account these corrections the relevant part of the K\"ahler potential receives the form~\cite{Antoniadis:2019rkh}
\be 
K=-2M_{P}^{2}\;\log({\cal V}+\xi_{0}+\eta_{0} \log{\cal V})+\cdots~, 
\label{Kahler1}
\ee 
where the dots stand for terms depending on the complex structure, $z_a$ and K\"ahler moduli $T_k=\tau_k+i a_k$. 
The general form of the  volume is given by ${\cal V}=\frac 16 \kappa_{ijk}t^it^jt^k$ where $t^i$ are the two-cycle K\"ahler moduli fields
and $\kappa_{ijk} $ are  triple intersection numbers on ${\cal X}$.
 In the subsequent, we work within a concrete scenario where the three K\"ahler moduli appear in the volume on equal footing and stabilize them through perturbative logarithmic loop corrections \cite{Antoniadis:2019doc}.  Following~\cite{Leontaris:2022rzj}, in particular,  we consider a framework that is based on a CY threefold corresponding to the polytope Id: 249 of the Kreuzer-Skarke~\cite{Kreuzer:2000xy}  CY database \cite{Altman:2014bfa}.  In this particular CY manifold there are  three K\"ahler moduli fields satisfying  the  simple relation $\tau_i=\mathfrak{a} t^jt^k$  between the two- and four-cycle moduli,
and $\mathfrak{a}$ is a positive constant associated with the intersection number. Then, the  volume is simply given by~\cite{Leontaris:2022rzj} 
\be 
\begin{split}
	\label{volume} 
{\cal V}&=\mathfrak{a} \,t^1t^2t^3 =\tau_1t^1= \tau_2t^2=\tau_3t^3 =\frac{1}{\sqrt{\mathfrak{a}}}\sqrt{\tau_1\tau_2\tau_3} ~.
\end{split}
\ee

 After the dimensional reduction,  the effective field theory model  could  be either some Grand Unified Theory (GUT) or directly the MSSM model  where the ordinary low energy (super)-fields 
  appear in appropriate representations of the EFT gauge group.
 Thus, in addition to the quantum corrections considered above, we also include matter fields in the K\"ahler potential.  These contributions are essential in studying soft supersymmetry breaking effects and cosmological inflation. Within the present context, in particular, we
focus on the Higgs sector which plays a vital role in implementing the scenario of hybrid inflation and investigating the possible production of dark radiation. 
Thus, we consider a generic set of  Higgs pairs $\Phi_i,\Phi_j$ 
  which are assumed to break the gauge group at some GUT scale
  much lower that the Planck scale $M_{P}$. We have also introduced a field  $S$, representing a gauge singlet superfield which realizes the trilinear superpotential couplings of the form  $S\Phi_{1}\Phi_{2}$. Such singlet fields are ubiquitous in effective string theory models. 
  In the setup described so far, the relevant terms of the superpotential have the following generic form 
  \begin{equation}
  	W= W_0+\kappa S(\Phi_{1}\Phi_{2}-M^2)+\cdots~,
  \label{superpotential}
  \end{equation}
  where  $\kappa$ is a coupling constant coefficient, $M$ is a high scale mass parameter whose values are below the string scale depending on which scale the Higgs field obtains  VEV, $W_{0}$ is the flux induced part introduced previously,  and dots stand for possible terms irrelevant to our discussion. 
  
{In our study, the procedure for working with the K\"ahler metric and incorporating matter fields can be compared to the study by \cite{Blumenhagen:2009gk}. This comparison extends to the examination of chiral matter, which is localized on magnetized D7-branes, and the more comprehensively understood fractional D3-branes found at singularities. When considering the second term within our K\"ahler metric, we explore soft terms linked to massless open strings concentrated at the intersections between $D3$ and $D7$ branes. However, it's important to note that the presence of such massless string states cannot be derived from the Dirac-Born-Infeld (DBI) or Chern-Simons (CS) actions alone.}\par
For the case of hybrid inflation within the framework proposed in our work, it becomes imperative to account for the contributions of scalar fields along with their fermionic superpartners in the K\"ahler potential. These contributions typically exhibit a generic dependence, expected to be in the form of $\tilde K_{ij} \Phi_i\bar{\Phi}_j$. Here, $\tilde K_{ij}$ represents a function of the moduli $\tau_k=(T_k+\bar T_k)/2$ and $S$. The precise characterization of $\tilde K_{ij}$ becomes clear upon identifying the origin of zero modes and in the context of type IIB theory, various possibilities arise (see for example~\cite{Aparicio:2008wh}).\par
{In the case of realistic constructions, chiral matter emerges from either the world volume of D7 brane stacks or at the intersections with other D7 branes. This is where gauge and scalar fields, such as $\Phi$ and $\bar\Phi$, manifest on the world volume by configuring D7 branes to wrap suitable divisors. The supermultiplets involving these scalar fields, $\Phi$ and $\bar\Phi$, exhibit a scaling dependence, with the leading contribution taking the form \cite{Lust:2004fi}:
\begin{equation}\nonumber
\frac{\Phi_i\bar\Phi_i}{T_k+\bar T_k} \rightarrow \frac{\Phi_i\bar\Phi_i}{{\cal V}^{2/3}}.
\end{equation}
Furthermore, it's worth mentioning that there are next-to-leading order $\alpha'$ contributions \cite{Blumenhagen:2009gk} of the form Re$(S)/(T_k+\bar T_k)$, however, for our specific purposes, these contributions remain insignificant and, hence, are disregarded.}
 
Including contributions of the above matter fields in  the K\"ahler potential~(\ref{Kahler1}), we obtain the following generic form  
 \begin{equation}\label{gk} 
 \begin{split}
 {\cal K} = -2 M_{P}^{2}\;\textrm{log} \left[\left(\prod_{k=1}^3\left(T_k+\bar{T_k}\right)\right)^{\frac 12}+{\cal C}\right]
 +\sum_{k=1}^3\dfrac{a_k}{T_k+\bar{T_k}}f_k(\Phi_i,\Phi_2..., S, \bar{S})
 \end{split}
 \end{equation}
 where ${\cal C}$ is a function with a logarithmic dependence on the three moduli fields $T_{1,2,3}$ defined as
  \begin{equation}
  {\cal C} = \xi_{0}+\eta_{0} \textrm{log} \left(\prod_{k=1}^3\left(T_k+\bar{T_k}\right)\right).
  \end{equation}
  In the above equation the parameter $\xi_0$  stands for $\alpha'^3$ corrections~\cite{Becker:2002nn} and is proportional to the Euler characteristic   $\chi_{CY}$ of the Calabi-Yau manifold
  \begin{equation}
  \xi_0= -\frac{\zeta(3)}{4} \chi_{CY}\,,
  \label{xi}
  \end{equation}
and $\eta_0$ is an order one coefficient~\cite{Antoniadis:2019rkh}.
 The functions $f_k$ in Eq. \eqref{gk} describe the visible Higgs sector.
For simplicity, we will assume that all $f_k$  are the same and have the following form
\begin{equation}\label{gk2}
	f(\Phi,\overline{\Phi},S) = \alpha\Phi_{1}\Phi_{1}^{\dagger}+\beta \Phi_{2} \Phi_{2}^{\dagger}+\gamma S S^{\dagger}+\lambda(\Phi_{1}\Phi_{2}+h.c.)~,
\end{equation}  
where $\alpha, \beta, \gamma$ and $\lambda$ are dimensionless couplings. Furthermore, we will adopt the approach used in \cite{Blumenhagen:2009gk} and we will express the matter contribution term in~(\ref{gk}) in terms of the compactification volume~(note that $ T\propto {\cal V}^{2/3}$). Regarding the origin of these fields, in the present work, we focus on matter fields residing on magnetized $ D7$-branes, as well as chiral fields generated at $D7$-brane intersections. Notice, however, that the general analysis also applies to states associated with the excitations of open strings stretching between $D7$-$D3$ branes or having both ends on the same $D3$ brane. The choice of the configuration dictates the modular weights in the K\"ahler potential. For our purposes, we consider the modular weights to be 1~\cite{Aparicio:2008wh}. 
Hence, in large volume compactifications,  the K\"ahler potential in Eq. \eqref{gk} takes the
form \cite{Aparicio:2008wh, Hebecker:2014gka, Conlon:2006tj}~\footnote{Notice that in our case this formula is identified with 
the expansion of the warped form of the K\"ahler potential 
	\[ K= -2\log \left((T+\bar T)^{3/2}+\xi+\eta\log{\cal V}- \frac{a}{2} (T+\bar T)^{1/2}\varphi_i\bar\varphi_i\right)
	= -2 \log \left({\cal V}+\xi+\eta\log{\cal V}- \frac{a}{2} {\cal V}^{1/3}\varphi_i\bar\varphi_i\right)\].}
\begin{equation}\label{kahlermp1}
{\cal K} = -2 \textrm{log} \left[\mathcal{V} + \xi_{0}+\eta_{0} \textrm{log} (\mathcal{V})\right]+\frac{3a}{\mathcal{V}^{2/3}}\left[\alpha\Phi_{1}\Phi_{1}^{\dagger}+\beta \Phi_{2} \Phi_{2}^{\dagger}+\gamma S S^{\dagger}+\lambda(\Phi_{1}\Phi_{2}+h.c.)\right]
\end{equation}
where  $a$ is a dimensionless constant and  $\mathcal{V}$ is defined in~(\ref{volume}). Note that in Eq. \eqref{kahlermp1} and from now on we adopt $M_{P}=1$ units.

Up to this point, we have presented the minimum number of moduli and matter fields that are necessary for our subsequent analysis. Next, we proceed with the computation of the scalar potential which is essential to investigate the properties of the model and compute the various cosmological and other phenomenological observables.

\section{The effective potential}\label{sec3}

The scalar potential of the effective field theory model receives various contributions. As we will see shortly, in the present construction there are F- and D-terms associated with the moduli sector, and  contributions 
from the EFT matter fields, as well as supersymmetry breaking terms. 
We start with the F-term potential  which  is given by the generic formula 
\be
V_{F}=e^G\left(G_iG_{i j^*}^{-1}G_{j^*}-3\right)~,\label{VGK}
\ee 
where 
\[ G= \mathcal{K}+\log|W|^2\equiv  \mathcal{K}+\log W+\log W^*~,
\]
and the indices $i,j$ in \eqref{VGK}  denote the derivatives with respect to the various moduli and other fields.

 Computing the derivatives, and substituting in  Eq. \eqref{VGK} while keeping only the leading order terms, the F-term potential receives the following simplified form~\footnote{The complex field $S$ has the general form $S=\mid S\mid e^{i\theta}$ and we choose $\theta=0$ to align it with the real axis. So in Eq. \eqref{vt} and the subsequent analysis, $S$ denotes the real part of the field.}

\bea \label{vt}
V_{F}&\simeq & \frac{\kappa ^2  \alpha \beta\left(M^2-\varphi_{1} \varphi_{2}\right)^2+\gamma \kappa ^2  S^2 \left(\alpha  \varphi_{1}^2+\beta  \varphi_{2}^2\right)}{3 a  \alpha \beta \gamma  \mathcal{V}^{4/3}}+\frac{3 W_0^2 (2 \eta_{0} \log\mathcal{V}-8 \eta_{0}+\xi_{0} )}{2 \mathcal{V}^3} ,
\eea 
where  $\varphi_{1}$ and  $\varphi_{2}$ are the bosonic components of the superfields $\Phi_{1}$ and $\Phi_{2}$.

 \begin{figure}[t]
 	\centering \includegraphics[width=7.95cm]{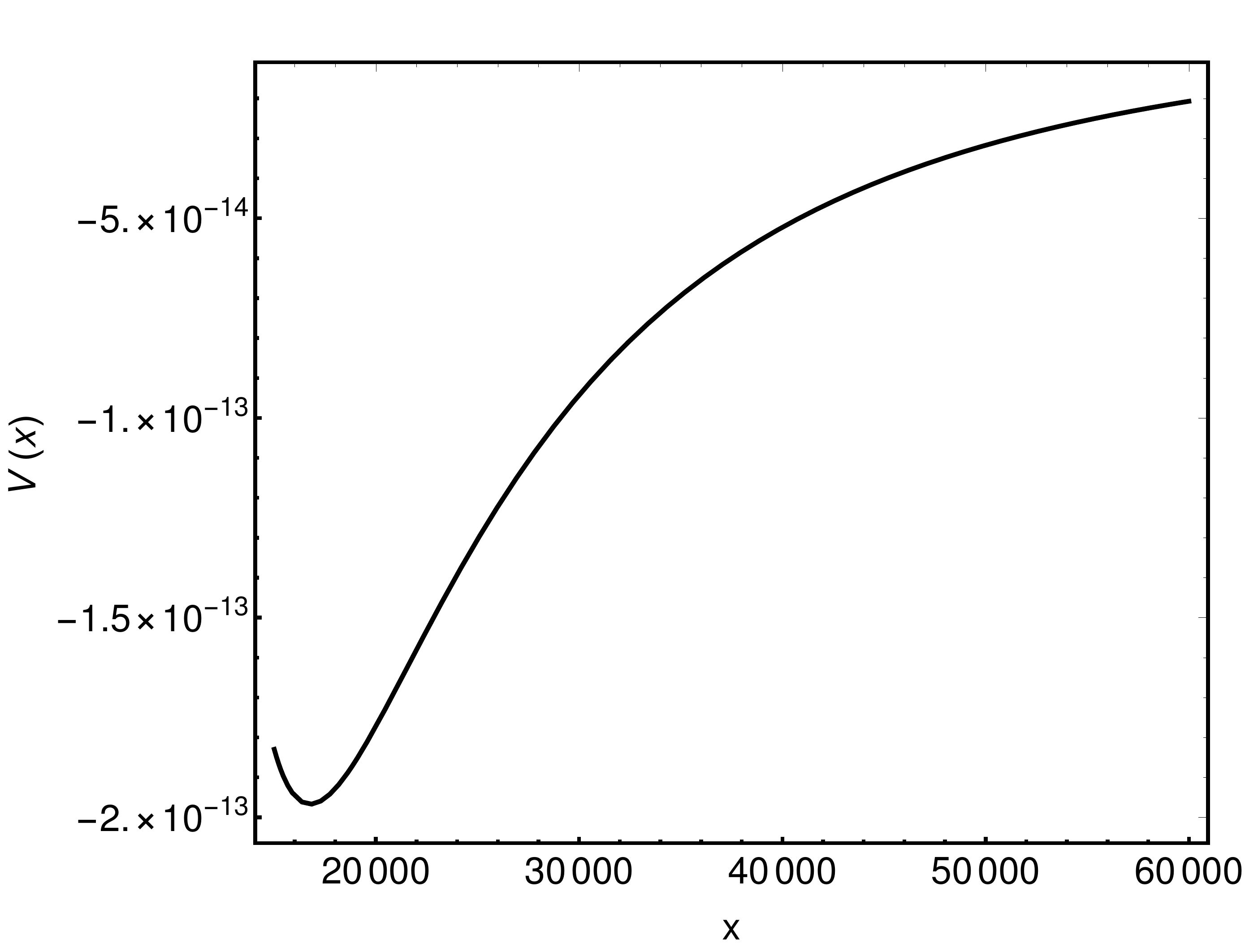}
 	\centering \includegraphics[width=7.7cm]{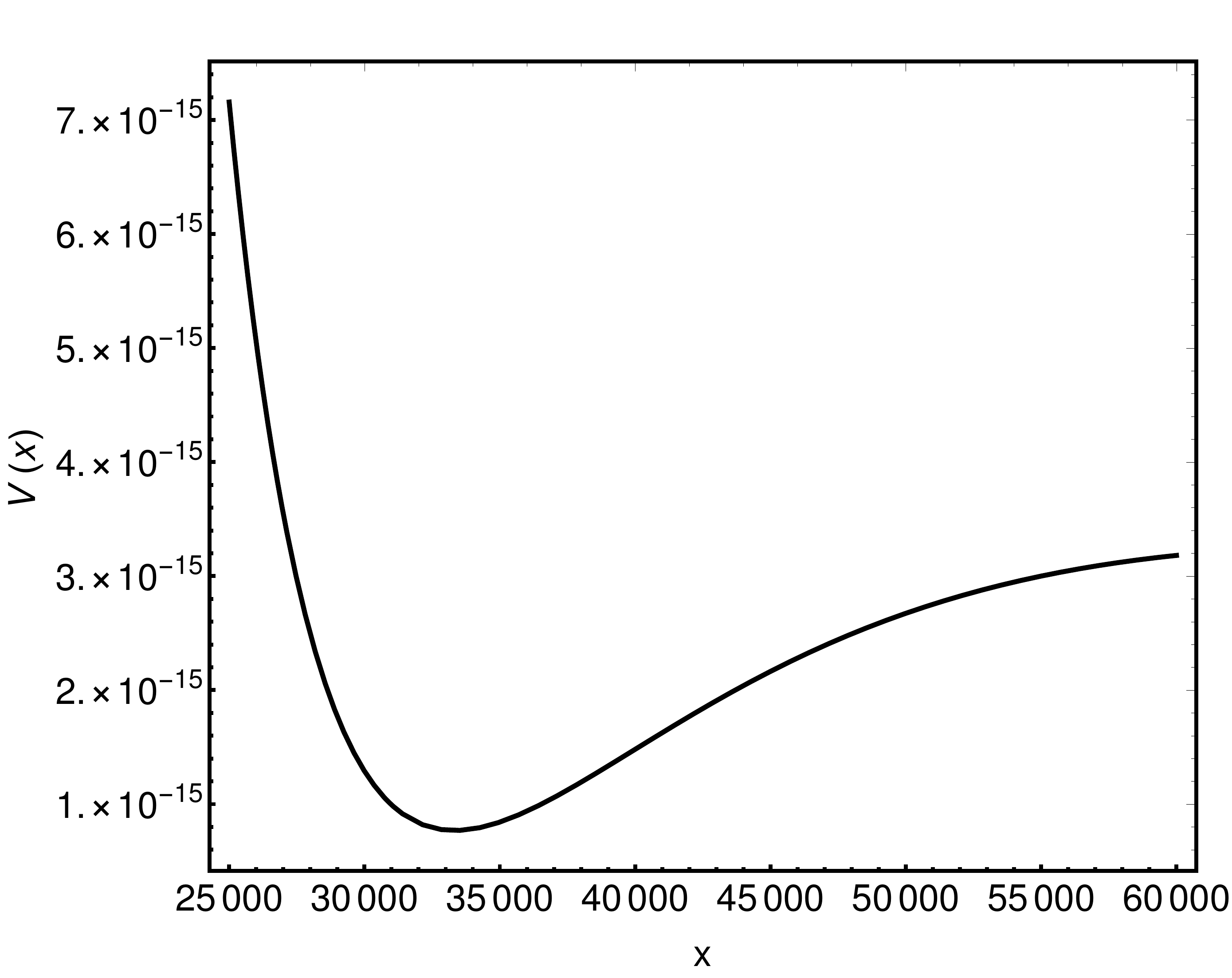}
 	\caption{\small Plots of the potential along the volume direction. The left panel shows the F-term potential, while the right panel also includes the D-term potential. We choose $\xi_{0}=10$, $\eta_{0}=-0.92$, $S=0$, $\varphi_{1,o}=\varphi_{2,o}=M$, $\kappa=0.1$ and $\gamma=1.$ Here $x$ represents the volume, $x\equiv\mathcal{V}$.}
 	\label{fig:pd}
 \end{figure} 
At the extrema of the F-term potential, the fields take the following values~\footnote{In more general EFT backgrounds it is possible that minimization with respect to the fields $S,\Phi_i$ leads to a potential of the form $V\sim \frac{a}{5 {\cal V}^{4/3}}+\frac{1}{3}\frac{b+\eta \log{\cal V}}{{\cal V}^3}$. In this case it is possible to have a dS minimum with the volume acquiring a value  ${\cal V}_{o}^{5/3}=\left(\frac{9n}{4a}\right)\,{\cal W}\left(\frac{4 a }{9 n}\left(e^{\frac{1}{3}-\frac{b}{n}}\right)^{5/3}\right)$, where ${\cal W}$ is the product-log (Lambert) function.}
\begin{equation}
S_o=0,\quad \varphi_{1,o} \varphi_{2,o}=M^{2},\quad \mathcal{V}_o=e^{\frac{13}{3}-\frac{\xi_{0} }{2 \eta_{0}}}.
\label{SPhimin}
\end{equation}
Substituting the solution~(\ref{SPhimin})  into~(\ref{vt})  we obtain
\begin{equation}\label{vfm}
V_F^{extr.}=\frac{3 |W_0|^2 \eta_{0}(2 \log\mathcal{V}_o-8 +\frac{\xi_{0}}{\eta_{0}} )}{2 \mathcal{V}_o^3} \equiv 
\eta_{0}\,\frac{|W_0|^2 }{ \mathcal{V}_o^3}\,\cdot 
\end{equation}
A simple analysis shows that this is a minimum of the potential as long as   $\eta_0<0$. 
However, since $\frac{|W_0|^2}{{\cal V}_{o}^{3}}>0$, the potential~(\ref{vfm}) at the minimum acquires a negative value,
 it turns out that the F-term potential predicts an anti-de Sitter (AdS) vacuum.
In Fig. \ref{fig:pd}, left panel, the F-term potential
with AdS minimum is plotted for a choice of the parameters $\xi_0, \eta_0, W_0$.

Despite the forgoing negative F-term contribution, the potential can be lifted to a de Sitter minimum, once   D-term contributions ~\cite{Burgess:2003ic, Haack:2006cy} associated with the universal $U(1)$  symmetries of the D7-branes are taken into account~\cite{Antoniadis:2018hqy}. {In this context, the D-term potential is generated in hidden sectors by D7-branes wrapping the 3 divisors whose volumes are $\tau_i$}. More precisely, in the present geometric setting, there are  D-term contributions due to the universal $U(1)$ factors associated with the $D7$ brane stacks. These terms   have the general form~\cite{Antoniadis:2018hqy,Haack:2006cy}
\be 
\textcolor{black}{V_D= \sum_{i=1}^3\frac{g_{D7_i}^{2}}{2}\left(Q_j\partial_{T_j} K+\sum_{j\neq i} q^{j}_{i} |\Theta^{j}_{i}|^2\right)^{2} ,\label{VD7} }
\ee 
where ${g^{2}_{D7_i}}=({\rm Re}T_i)^{-1} $ and $Q_j,q^{j}_{i}$ are ``charges'', while {$\Theta^{j}_{i}$} represent matter fields charged
under the $U(1)$ gauge factors. Some aspects regarding strings
at $D7$ intersections in particular have been recently discussed in \cite{Antoniadis:2021lhi}. The fields {$\Theta^{j}_{i}$} carry charges under the $U(1)$ factors associated with the D-branes.  A compelling question then arises regarding the possible contributions of these fields in the D-term, and hence, in the scalar potential. One possible scenario is that non-zero field vevs are chosen so that the D-term vanishes, and in this case, the uplift should be realized by a suitable modification (see footnote 10), or the standard procedure of introducing $\overline{D3}$  branes (see \cite{Kachru:2003aw}).
Here we make a simplified assumption that these singlets have vanishing vevs. Whether the vevs of these fields are zero or not,   depends on the specific details of the effective model. One possibility is to assume that in the effective field theory limit, all these fields {$\Theta^{j}_{i}$}  are minimized at $\langle \Theta^{j}_{i}\rangle =0$. Alternatively, even in the case of non-zero vevs, however, it might happen that there could be accidental cancellations that can reduce the significance of their contributions.  
Some related discussions regarding these issues can also be found in~\cite{Achucarro:2006zf}.
{Here, since the D-term potential is only used to uplift the non-supersymmetric AdS vacuum of the F-term potential, following closely previous works (see for example \cite{Antoniadis:2018hqy}, \cite{Burgess:2003ic} and \cite{Haack:2006cy}) we make the reasonable assumption that
moduli fields dominate over other fields vevs.  Therefore,  it is adequate for our purposes
to assume that the flux induced D-term piece $\propto Q_j\partial_{T_j}K$ dominates, and therefore we minimize the potential by setting  $\langle \Theta^{j}_{i}\rangle$ vevs zero. 
For  $\langle\Theta^{j}_{i}\rangle=0 $ the second term in the parenthesis vanishes 
and each component of the D-term acquires a simple -model independent- form $V_{D_i}\approx  Q_i^2/\tau_i^3$. 
Then, the total  D-term potential,  being the sum of three components, is  approximated by~\cite{Antoniadis:2018hqy}
\be 
{V}_{ \cal D}\; = \;\sum_{i=1}^3\frac{d_i}{\tau_i} \left(\frac{\partial { {\cal K}}}{\partial {\tau_i}}\right)^2
\;\approx \;
\sum_{i=1}^3 \frac{d_i}{\tau_i^3}\;\equiv\; \frac{d_1}{\tau_1^3}+\frac{d_3}{\tau_3^3}+\frac{d_2\tau_1^3\tau_3^3}{{\cal V}^6}~,\label{VDappr}
\ee 
Here,  $d_i$ are positive constants related to the charges $d_i\sim Q_i^2>0$. {{\color{black} In our computations, we focused solely on the simplest scenario involving diagonal terms. However, in more general cases, cross-terms are also present. We observed that these cross terms exhibit the same order of magnitude in relation to the expression of $\tau_i$'s. Numerical analysis reveals the existence of minima in dS, albeit with slightly different coefficient values, $d_i$}}. Furthermore, using the volume formula 
${\cal V}^2=\tau_1\tau_2\tau_3$, we have  substituted  the modulus $\tau_2$ with its equivalent $\tau_2={\cal V}^2/(\tau_1 \tau_3)$ . 
Then, the total effective potential is the sum  $V_{\rm eff}= V_F+V_D$  which can be minimized with respect to
$\mathcal{V}$ and the two remaining K\"ahler moduli $\tau_{1,2}$.  We have assumed that the F-part of the potential
depends on the total volume, hence the explicit dependence of $V_{\rm eff}$ on $\tau_1,\tau_3$ only comes through the  $V_D$ part.  It is found that the 
two minimization conditions with respect to $\tau_{1,3}$ determine the ratios between the moduli, i.e., $\left(\frac{\tau_i}{\tau_j}\right)^3=\frac{d_i}{d_j}$~\cite{Antoniadis:2020stf}.
Expressed in terms of the stabilized total volume $\mathcal{V}$, the conditions for the two $\tau_i$ can be written as
\[\tau_i^3 = \left(\frac{d_i^2}{d_kd_j}\right)^{\frac 13}{\mathcal{V}}^2~,  \]
where $i=1,3$. In this case, the D-term potential receives the simple form
\be 
V_{D}\approx  \frac{d}{\mathcal{V}^2}~, \quad \text{with } \qquad  d=3(d_1d_2d_3)^{\frac 13}.\;
\label{Dterm_final}
\ee 
{\color{black} {At tree-level, the potential exhibits two flat directions when the variable S equals zero. In this scenario, one can choose different values for the variables $\phi_1$ and $\phi_2$ to create distinct minima in the potential. However, it's essential to note that this principle applies solely when considering the F-term potential. The total potential at tree level comprises both F-term and D-term components. The D-term potential encompasses two parts: one stemming from the moduli and the other from the matter fields. To nullify the contribution from the matter fields in the D-term potential, we utilize D-flat directions where 
  $\varphi_{1}=\varphi_{2}=\varphi$ and  $\alpha=\beta$}}. Then the effective potential  has the following form
\bea \label{eq1}
V_{\text{eff}}&\simeq & \frac{\kappa ^2  \alpha\left(M^2-\varphi^{2}\right)^2+2\gamma \kappa ^2  S^2 \varphi^{2}}{3 a  \alpha \gamma  \mathcal{V}^{4/3}}+\frac{3 W_0^2 (2 \eta_{0} \log (\mathcal{V})-8 \eta_{0}+\xi_{0} )}{2 \mathcal{V}^3}+\frac{d}{\mathcal{V}^2}~\cdot 
\eea 
 \begin{figure}[t]
 	\centering \includegraphics[width=11cm]{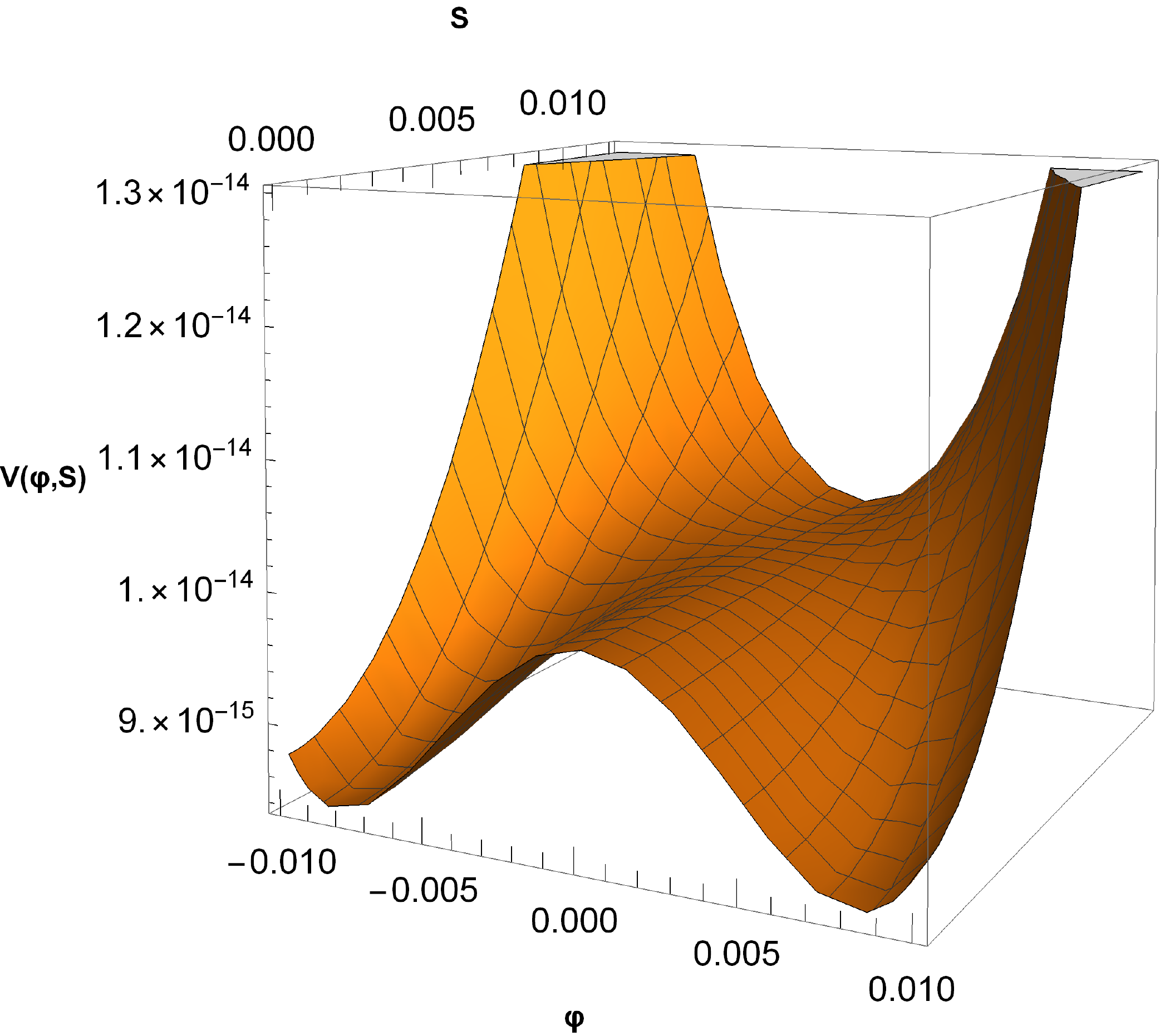}
 	\caption{\small The  shape of the effective potential in $\varphi-S$ plane for the choice of parameters  $\xi_{0}=10$, $\eta_{0}=-0.92$, $\mathcal{V}_{0}=32000$, $\kappa=0.1$, $\gamma=1.$ and $d=10^{-4.52}$.}
 	\label{fig:pd3}
 \end{figure}
The shape of the potential along the volume modulus ${\cal V}$ when both F- and D-terms are included is shown in the right panel of Fig.~\ref{fig:pd}. As can be seen, a positive D-term is sufficient to uplift the potential along the volume direction so that we achieve a de Sitter minimum. In order to find the extrema of the potential along the $\varphi$ and $S$ directions, we require the vanishing of its corresponding derivatives. Thus, for  $\varphi$ we impose the condition
\bea\label{eqphi}
\dfrac{dV_{\text{eff}}}{d\varphi}&=&0\Rightarrow\frac{\kappa ^2 \left(4 \alpha  \varphi ^3-4 \alpha  M^2 \varphi +4  \gamma  S^2 \varphi \right)}{3 a \alpha \gamma  \mathcal{V}^{4/3}}=0~,
\eea
which is solved for the following three $\varphi$-values 
\bea\label{eqphi1}
\varphi=0,\quad \varphi_{\pm}=\pm \sqrt{\  M^2-\dfrac{\gamma }{\alpha} S^2}.
\eea
Similarly along the $S$ direction
\bea \label{eqs1}
\dfrac{dV_\text{eff}}{dS}&=&0\Rightarrow\frac{4 \kappa ^2 S \varphi ^2}{3 a \alpha  \mathcal{V}^{4/3}}=0~,
\eea
which yields 
\bea\label{eqs11}
S=0.
\eea

Combining Equations \eqref{eqphi1} and \eqref{eqs11}, in the large volume limit we obtain the following solutions
\bea
(S=0,\varphi=0),\quad (S=0,\varphi=\pm M). 
\eea

We have already dealt with the minimization of $V_F$  with respect to the volume modulus. However, 
in the presence of D-terms, the minima along the volume direction are shifted. Thus, requiring the vanishing of  the derivative of (\ref{eq1}) with respect to ${\cal V}$, we obtain the equation
\bea
\dfrac{dV_{\rm eff}}{d{\cal V}}=0&\Rightarrow&
-\frac{4 \kappa ^2 \left(\alpha \varphi ^4+\alpha  M^4-2 \alpha  M^2 \varphi ^2+2   \gamma  S^2 \varphi ^2\right)}{9 a \alpha  \gamma  {\cal V}^{7/3}}-\frac{2 d}{{\cal V}^3}\nonumber
\\
 &-&\frac{9 \left(2 \eta_{0} W_0^2 \log ({\cal V})-8 \eta_{0} W_0^2+\xi_{0} W_0^2\right)}{2 {\cal V}^4}+\frac{3 \eta_{0} W_0^2}{{\cal V}^4}=0~.
\eea
In the large volume limit, to a good approximation,  the above equation gives the solution
\bea
{\cal V}_o\approx \frac{9  \eta_{0} W_0^2 }{2 d}\mathcal{W}\left(\frac{2 d e^{\frac{13}{3}-\frac{\xi_{0} }{2 \eta_{0}}}}{9  \eta_{0} W_0^2}\right),
\eea
where $\mathcal{W}$  represents the product-log (Lambert) function. 
 The shape of the scalar potential $V_{\rm eff}$ in the $\varphi$-$S$ plane is displayed in  Fig. \ref{fig:pd3}. As discussed earlier the D-term contribution in the effective potential ensures the existence of de Sitter vacua when $S$ approaches zero.  

\subsection{Inflationary phase}

Up to this point, we have analyzed in detail the scalar potential of the effective theory and the role of the various fields in its final shape. Therefore we are now fully equipped with all the tools and the necessary ingredients to examine whether cosmological inflation is realized in the present model.  In a previous  approach, within the same  type-IIB framework and  
the geometric setup of intersecting D7 brane stacks,  the inflaton field was associated with the logarithm of the compactification volume modulus.
The accumulation of the required  60 efolds
to realize slow-roll inflation was converted to a lower bound on the minimum vacuum energy~\cite{Antoniadis:2020stf},  yet much bigger 
than the cosmological constant.   Subsequently,  a new ``waterfall'' field was introduced which adds a new 
direction of the potential.  The waterfall field rolls down towards the new lower minimum while at the same time, it ends inflation.
It was shown~\cite{Antoniadis:2021lhi} that this role can be realized by open string states oscillating near the intersections of the D7 stacks.

In the present case where the physical states from the effective
theory model have been taken into account,  new possibilities have emerged.  At the minimum along the compactification volume modulus ${\cal V}$, the Higgs field $\varphi$, and the singlet $S$ add new directions (transverse to that of ${\cal V}$)  providing new lower minima for the scalar potential.  As such, they are potential candidates for waterfall fields. In the present scenario inflation  proceeds along the local minimum $\varphi = 0$
 (the inflationary track), beginning at large $S$ values. An instability occurs at the waterfall point $S_{c}^{2}  = M^2$ , which is the value of $S$, such that $S_c=\frac{\partial^{2}V}{\partial S^{2}}\mid_S=0$. At this point, the field falls naturally into one
of the two minima at $\varphi=\pm M$. At large $S$, the scalar potential is approximately quadratic in $\varphi$, whereas at $S=0$, equation \eqref{eq1} becomes a Higgs potential. Along the inflationary track the constant term
 $$V_0^{vol}=\frac{\kappa ^2 M^4}{3 a  {\cal V}_{o}^{4/3}}+\frac{3 W_0^2 (2 \eta_{0} \log ({\cal V}_{o})-8 \eta_{0}+\xi_{0} )}{2 {\cal V}_{o}^3}~,$$
 is present at tree level, thus SUSY is broken during inflation. When SUSY breaks, splitting between fermionic and bosonic mass multiplets is created and contributions to radiative corrections occur. Following  \cite{Brignole:1993dj, Conlon:2006wz}, the soft terms are\footnote{\textcolor{black}{Note that the factor ${\cal{V}}_{0}^{2/3}/(3a\gamma)$ originates from the transition to the canonically normalized field $s= S \sqrt{3a\gamma}/ V_{0}^{1/3}$. See also Appendix \ref{AppA}.}}
 \begin{equation}
\Delta V_{\text{soft}}=\left[(m^{2}_{3/2}+ V_{0})-\frac{2}{3{\cal V}_{o}^{2}}W_o^2+\cdots\right]M^2 y^2{\frac{{\cal V}_{o}^{2/3}}{3 a \gamma}}=M_{s_c}^2y^2~,
 \end{equation}
 where in the above equation, $y$  defines the ratio $y=s/M$, and $$M_{s_{c}}=\left[(m^{2}_{3/2}+ V_{0})-\frac{2}{3{\cal V}_{o}^{2}}W_o^2+\cdots\right]M^2 {\frac{{\cal V}_{o}^{2/3}}{3 a \gamma}}~,$$
  represents the soft mass parameter for canonical normalized field $y$, and $V_{0}$ is the minimum of the potential \eqref{eq1}.\footnote{The soft mass parameter, $M_{s_{c}}$, depends on both the gravitino mass, $m_{3/2}$, and the constant parameter $W_{0}$. By choosing $W_{0}$ to be very small, the dominant term of the soft masses becomes dependent on the gravitino mass \cite{Demirtas:2020ffz}.} For  appropriate set of parameters, $V_{0}$ takes the value of the cosmological constant. The first extremum $(s=0,\varphi=0)$ is the maximum of the potential. The $\varphi=0$ corresponds to the trajectory of the standard hybrid inflation for which $\{\varphi = 0$,\,$s > M\}$.  When the inflaton reaches $s=M$, the waterfall field takes over and the inflaton moves towards the minimum at $\varphi = \pm M$. Moreover, SUSY is broken along the inflationary track, and the radiative corrections 
 along with the soft SUSY-breaking potential $V_{\text{soft}}$ can lift its flatness, while also providing the necessary slope for driving inflation. The effective contribution of the one-loop radiative corrections can be calculated using the Coleman-Weinberg formula \cite{Coleman-Weinberg: 1973} as 
  \begin{equation}
	\Delta V_{\text{1-loop}}=  \frac{\kappa ^4 M^4y^4}{144\pi ^2 a^2  \gamma^2 {\cal V}_o^{4/3} }\left[F(y)-\left(\frac{ 1}{54 a^2 \alpha^2 \gamma^2}-\frac{3}{2}{\cal V}_o^{8/3}\right)\right]~,
\end{equation}
 where
 \begin{equation}
 F(y)=\frac{1}{81a^2\alpha^2 \gamma^2}\log \left(\frac{\kappa ^2 M^2 y^2}{27 a^2 \alpha\gamma^2  Q^2 {\cal V}_o^{2/3}} \right)-{\cal V}_o^{8/3}\log\left(\frac{\kappa ^2 M^2 y^2 {\cal V}_o^{2/3}}{3 a \gamma Q^2}\right)~.
 \label{logcw}
\end{equation}
It should be observed that the Coleman-Weinberg correction is computed along the inflationary trajectory, wherein the field $s$ takes a non-zero value while the field $\varphi$ remains fixed at zero. Furthermore, we set $V_{o}=3.2\times 10^4$ for the volume field\footnote{\textcolor{black}{For a discussion on multifield inflation scenario in this type of models see  \cite{Antoniadis:2018ngr}.}}, and the mass spectrum solely depends on $s$. The detailed calculations are provided in Appendix \ref{AppB}.

Including the various contributions computed above, we may write the scalar potential along the inflationary trajectory (i.e. $\varphi_{1}=\varphi_{2}=0$) as follows,
\begin{equation}
\begin{gathered}
V \simeq V_{F} + V_{D} + \Delta V_{\text{1-loop}} + \Delta V_{\text{soft}}, \\
\simeq \kappa ^{2}M^{4}\left( \frac{V_0^{vol}}{\kappa ^{2}M^{4}} + \frac{\kappa^2 y^4 F(y)}{144 \pi ^2 a^2 \gamma ^2 \mathcal{V}_{o}^{4/3}} - \frac{\kappa ^2 y^4}{144 \pi ^2 a^2 \gamma ^2 \mathcal{V}_{o}^{4/3}}\left(\frac{1}{54 a^2 \alpha^2 \gamma^2} - \frac{3}{2}{\cal V}_o^{8/3}\right) + \frac{M_{s_{c}}^2 y^2}{\kappa ^2 M^2 }\right)~,
\label{scalarpot}
\end{gathered}
\end{equation}

\noindent $M_{s_{c}}$ is the soft-mass parameter of the field $s$, and $V_{0}\approx 0$ due to its extremely small magnitude compared to $m^{2}_{3/2}$ . 
 
 The predictions for the various inflationary observables  are estimated using the standard slow-roll parameters defined as,
 \begin{align}
 \epsilon &= \frac{1}{2}\left( \frac{1}{M}\right)^2
 \left( \frac{V'}{V}\right)^2, \,\,\,
 \eta =  \frac{1}{M^{2}}
 \left( \frac{V''}{V} \right), \notag\\
 \xi^2 &= \frac{1}{M^4}
 \left( \frac{V' V'''}{V^2}\right),
 \label{slowroll}
 \end{align}
 where prime denotes the derivative with respect to $y$. \textcolor{black}{Note that, the presence of the mass parameter $M$ in \eqref{slowroll} comes from the definition of the field $y$ as the ratio $y=s/M$}.  In the slow-roll approximation, the scalar spectral index $n_s$, the tensor-to-scalar ratio $r$ and the running of the scalar spectral index $\alpha_{s}\equiv dn_s / d \ln k$ are given by,
 \begin{align} \label{nsr}
 n_s&\simeq 1+2\,\eta-4\,\epsilon, \,\,\,\,\,\,\,\,\,\,\,
 r \simeq 16\,\epsilon, \notag\\
 \alpha_{s} &\simeq 16\,\epsilon\,\eta
 -24\,\epsilon^2 - 2\,\xi^2.
 \end{align}
 The value of the scalar spectral index $n_s$ in the $\Lambda$CDM model is $n_s = 0.9665 \pm 0.0038$ \cite{Planck:2018jri}.
 
 The amplitude of the scalar power spectrum is given by,
 \begin{align}
 A_{s}(k_0) = \frac{1}{24\,\pi^2\,}
 \left( \frac{V(y_0)}{\epsilon(y_0)}\right),  \label{curv}
 \end{align}
 where  $A_{s}(k_0) = 2.137 \times 10^{-9}$ at the pivot scale $k_0 = 0.05\, \rm{Mpc}^{-1}$ as measured by Planck 2018 \cite{Planck:2018jri}.
 The relevant number of e-folds, $N_0$, before the end of inflation is,
 \begin{align}\label{Ngen}
 N_0 = 2M^{2}\int_{y_e}^{y_{0}}\left( \frac{V}{%
 	V'}\right) dy,
 \end{align}
 where $y_0 \equiv y(k_0)$ is the field value at the pivot scale $k_0$, and
 $y_e$ is the field value at the end of inflation. 
 As the case may be, the value of $y_e$ is fixed either by the breakdown of the slow-roll approximation ($\eta(y_e)=-1$) or by a `waterfall' destabilization occurring at the value $y_e = 1$. 
 \begin{figure}[t]
 	\centering \includegraphics[width=13cm]{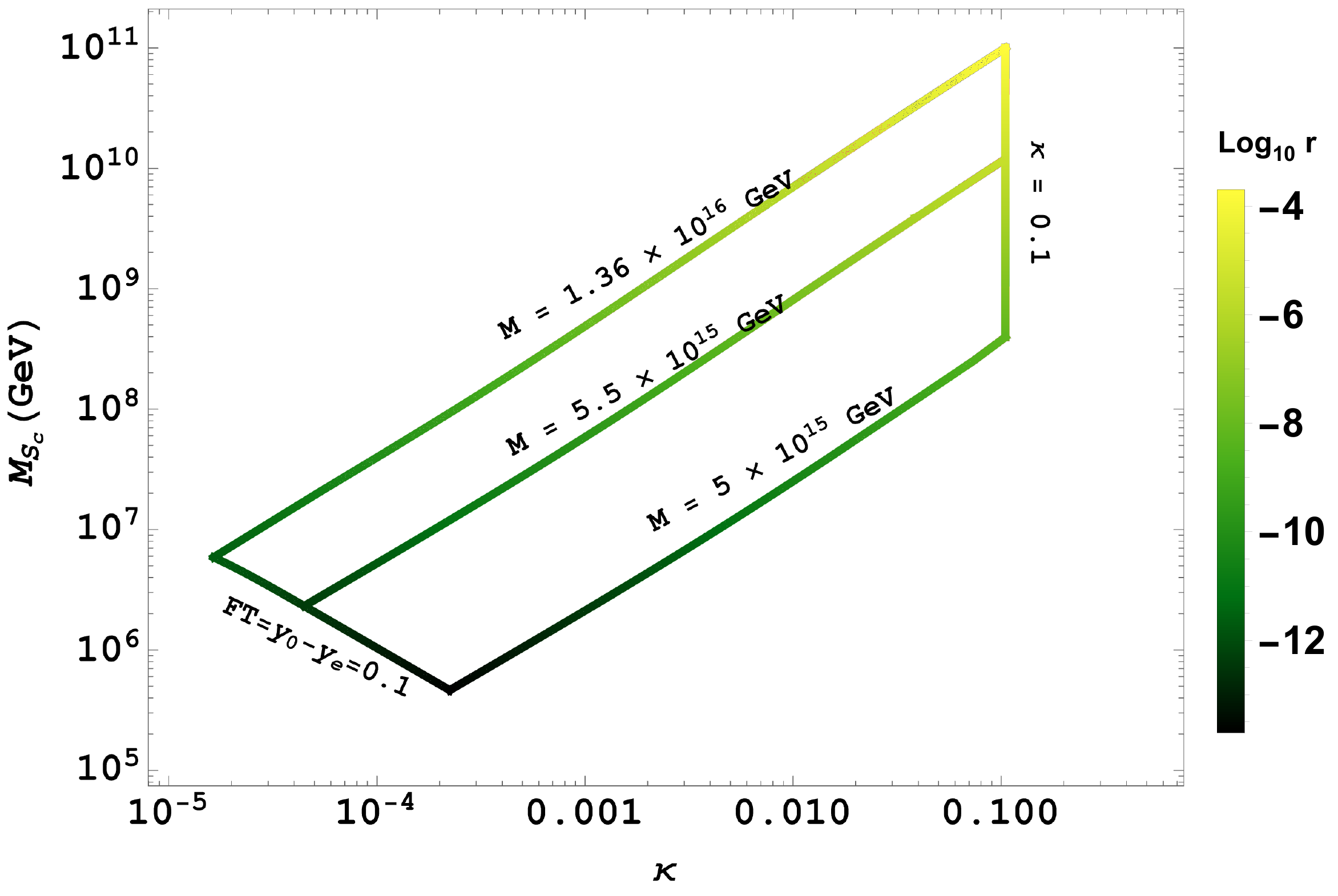}
 	\caption{\small Variations of $r$ and $M$ in $\kappa-M_{s_{c}}$ plane. The upper line has $M$ fixed at a value equal to $M_{string}$. The other two lines correspond to lower $M$ values, as indicated in the plot.}
 	\label{fig:results1}
 \end{figure} 
 
\subsection{Numerical results}

{The results of our numerical calculations are displayed in Fig. \ref{fig:results1}, which show the ranges of $r$, $M$ in the  \textcolor{black}{$\kappa-M_{s_{c}}$} plane. We consider up to second-order approximation on the slow-roll parameters, and for simplicity, we set $\gamma = 1$, $\alpha = 10^{-7}$, $\mathcal{V}_{0}=32000$ and $y_e = 1$. Moreover, we have fixed the spectral index $n_s$ to the central value ($n_s=0.96655$) from Planck's data. }

{We further require $\kappa\leq 0.1$, the Higgs mass parameter $10^{15}\leq M\leq M_{string}\sim 1/{\mathcal{V}_{o}}^{1/2}=5.5\times10^{-3}M_{p}=1.36\times 10^{16}$ GeV, and FT, which is defined as the difference of field value at pivot scale at the end of inflation $FT\equiv y_0-y_e\lesssim 0.1$. These constraints appear in Fig. \ref{fig:results1} as the boundaries of the allowed region in the \textcolor{black}{$\kappa-M_{s_{c}}$} plane. In our analysis, the soft SUSY contributions, along with the radiative corrections, play the dominant role in obtaining a parametric space consistent with experimental bounds, parametrized by \textcolor{black}{$M_{s_{c}}$} and $a$. Additionally, in the entire parameter space, $a$  remains in close proximity to the value of 1.}

{In our analysis, we carefully selected a parameter space where the contribution of the logarithmic terms of in (\ref{logcw}) is suppressed. However, for $\kappa\geq 0.1$, the logarithmic terms gradually become more dominant compared to the other components of the potential in (\ref{scalarpot}). This dominance results in a deviation of the spectral index ($n_{s}$) from the Planck bound. }

{For the scalar spectral index $n_s$ fixed at Planck's central value, $n_s=0.96655$, according to our numerical analysis, the exact range of parameters where acceptable solutions occur is presented below
 \begin{gather}
 \nonumber
 1.6 \times 10^{-5} \lesssim \kappa \lesssim 0.1, \\ \nonumber
 (1 \times 10^{15} \lesssim M \lesssim 1.3 \times 10^{16}) ~ \text{GeV}, \\ \nonumber
 (4.6\times 10^{5} \lesssim {M_{s_{c}}}\lesssim 1 \times 10^{11}) ~ \text{GeV}, \\ \nonumber
 1.9 \times 10^{-14} \lesssim r  \lesssim 1.4\times 10^{-4},\\\nonumber
 0.86  \lesssim a  \lesssim 1.
 \end{gather}
Examining the exact ranges of solutions, our calculations predict a low tensor-to-scalar ratio ($r$) compared to current experimental bounds. However, ongoing and future gravity waves experiments are expected to reach much smaller ranges of tensor-to-scalar ratio, comparable to our numerical predictions. Note that  Fig. \ref{fig:results1}  shows the parametric space where $0.86  \lesssim a  \lesssim 1$, aligning with the conditions for dark radiation, as discussed in the following section.}

\section{Reheating and dark radiation} \label{sec4}

After the end of inflation, the lightest moduli will begin to oscillate around their minima, acquiring a large energy density in the process. The decay products of the modulus fall into two categories.  The first are decays that go to the visible sector that is, particles of the SM, or its extensions such as the MSSM. The decays to visible matter induce reheating,  after which the standard hot  Big  Bang cosmology follows. In addition, there may also be decays to hidden sector states. The hidden sector contains several candidates for dark radiation, such as massless axions or light hidden gauge bosons. Let us consider the three K\"ahler moduli case, $T_k  =\tau_k+i a_k, {\cal V}=\sqrt{\tau_1 \tau_2\tau_3}$ where $a_k$ is the RR-axion. Decay  to  the  light  axion $a_k$ takes  place  primarily  through  the  supergravity kinetic terms for the K\"ahler moduli which read as,
\begin{equation} \label{Eq1}
\mathcal{L} \supset  K_{i\bar j} \partial_{\mu} T^i \partial^{\mu} T^{\bar j}.
\end{equation}
The tree level K\"ahler potential is,
\begin{equation}
K=-2\log\sqrt{(T_1+\bar T_1)(T_2+\bar T_2)(T_3+\bar T_3)} \, =\,-\log ({\tau_1 \tau_2\tau_3})+\cdots ~,
\end{equation}
where the dots represent constant terms that are ignored. 
Then, the K\"ahler  matrix is found to be
\begin{equation}
K_{i\bar j}=\frac{1}{4} {\rm diag}\left(\frac{1}{\tau_1^2},\frac{1}{\tau_2^2},\frac{1}{\tau_3^2}   \right)  .
\end{equation}
Therefore, Eq \eqref{Eq1} can be rewritten as \cite{Cicoli:2022uqa},
\[{\cal L}\supset   \frac{1}{4}\frac{1}{\tau_i^2}\partial_{\mu} \tau_i \partial^{\mu} \tau_{i} \;.\]
Note here that we have set the reduced Planck mass $M_{p}=1$. To put this into canonical form  we need to find the transformation
$\tau_i(u_i)$ so that
\[
{\cal L}\supset  \frac{1}{2}\sum_i \partial_{\mu} u_i \partial^{\mu} u_i \;.\]
This implies
\[ \tau_i= e^{\sqrt 2 u_i}\; .\]
The moduli fields for canonical kinetic terms take the form
\[  u_k =\frac{1}{\sqrt 2} \log \tau_k \;.\] 
The corresponding volume modulus is
\[  t = \frac{u_1+u_2+u_3}{\sqrt 3}= \frac{1}{\sqrt 3}\frac{1}{\sqrt 2}\sum_k \log \tau_k = \sqrt{\frac{2}{3}} \log{\cal V} \; .\] 
The transverse directions are
\[  u= \frac{u_1-u_2}{\sqrt 2}=\frac 12 \log{\frac{\tau_1}{\tau_2}},\; v=\frac{ u_1+u_2-2u_3}{\sqrt 6} =
\frac{1}{\sqrt {12}}\log{\frac{\tau_1\tau_2}{\tau_3^2}} \; .\]
Finally, reversing the above relations and expressed them in matrix form we have :
\[\left(
\begin{array}{c}
u_1 \\
u_2
\\
u_3 \\
\end{array}
\right)=
\left(
\begin{array}{ccc}
\frac{1}{\sqrt{3}} & \frac{1}{\sqrt{2}} & \frac{1}{\sqrt{6}} \\
\frac{1}{\sqrt{3}} & -\frac{1}{\sqrt{2}} & \frac{1}{\sqrt{6}} \\
\frac{1}{\sqrt{3}} & 0 & -\sqrt{\frac{2}{3}} \\
\end{array}
\right)\left(\begin{array}{c}
t \\
u\\
v \\
\end{array}
\right)\;. \]
{\color{black} Regarding the Lagrangian part for the axions we have \cite{Cicoli:2022uqa}:
  \begin{equation}\nonumber
    {\cal L}\supset - \frac{1}{4}\frac{1}{\tau_{i}^{2}}\partial_{\mu}a_{i}\partial^{\mu}a_{i}= -\frac{1}{4}e^{-2\sqrt{2}u_{i}}\partial_{\mu}a_{i}\partial^{\mu}a_{i} \;.
  \end{equation}
After expanding, we obtain the result
\begin{equation}\nonumber
   {\cal L}\supset  -\frac{1}{4}\left[\partial_{\mu}a_{i}\partial^{\mu}a_{i} - 2\sqrt{2}\left(u_{1}\partial_{\mu}a_{1}\partial^{\mu}a_{1}+u_{2}\partial_{\mu}a_{2}\partial^{\mu}a_{2}+u_{3}\partial_{\mu}a_{3}\partial^{\mu}a_{3} \right) \right] \;,
\end{equation}
where the first term is the pure kinetic energy term for axions, while the second corresponds to the interaction terms. Considering the interaction part and expressing the values $u_{1}$,$u_2$ and $u_3$ in terms of u, v, and t we have,
\bea
{\cal L}&\supset & \frac{1}{\sqrt 6} t \left(\partial_{\mu}a_1\partial^{\mu}a_1 +  \partial_{\mu}a_2\partial^{\mu}a_2+\partial_{\mu}a_3\partial^{\mu}a_3 \right) \nonumber\\
&&+ \frac{1}{2}u
\left(\partial_{\mu}a_1\partial^{\mu}a_1 - \partial_{\mu}a_2\partial^{\mu}a_2\right)\nonumber\\
&&+
\frac{1}{2\sqrt 3}v\left(\partial_{\mu}a_1\partial^{\mu}a_1 +  \partial_{\mu}a_2\partial^{\mu}a_2-2\partial_{\mu}a_3\partial^{\mu}a_3 \right).
\eea}
The decay rate of the lightest modulus  $u$ to axions is given by
\begin{equation}
\Gamma(u\to a_1a_1) =\frac{1}{64\pi} m_{u}^3 \;,
\label{gammau}
\end{equation}  
where $m_{u}$ is the modulus mass.  In the large volume scenario, a distinct hierarchy of mass scales is generated. \textcolor{black}{Here, we employ K\"ahler log corrections in stabilizing the moduli
fields, yet the overall spectrum remains consistent with the findings
discussed in }\cite{Cicoli:2011yy}. The mass eigenstates after diagonalization in units of $M_p=1$ can be written as,
\begin{equation*}
m_{t}=m_{u}=m_{v}\sim \frac{1}{\mathcal{V}_{o}^{3/2}},\quad m_{a_i}\sim e^{-2\pi \mathcal{V}_{o}^{2/3}}, \quad  m_{soft}\sim \frac{1}{\mathcal{V}_{o}^{2/3}}, \quad 
m_{3/2}\sim \frac{1}{\mathcal{V}_{o}}, 
\quad M_{string} \sim \frac{1}{\mathcal{V}_{o}^{1/2}}.
\end{equation*}

 Similarly, the dominant visible-sector decay channel is the decay to Higgs bosons. 
 Specializing to the MSSM case,  we can make the identifications $\Phi_{1}=H_u$ and $\Phi_{2}=H_d$. Each of these is a two-component complex field,  so there are eight degrees of freedom. Then the decay rate can be derived by including the matter contribution to the K\"ahler potential
\begin{equation}
{\cal L} \supset \frac{3 a \lambda}{\sqrt{2}} H_u H_d  \square u +h.c. +\cdots~.
\end{equation}
The dominant contribution to the decay of light moduli $u$ comes from the Giudice-Masiero coupling~\cite{Giudice:1988yz} 
 $3 a \lambda H_u H_d  \square u$ as all other couplings are mass-suppressed \cite{Angus:2013zfa}. Each field is a complex doublet, hence we include the partial widths from each of the four decay channels.  This yields
\begin{equation}
\Gamma(u\longrightarrow H_u H_d)=\frac{9 a^2 \lambda^2}{8\pi}m_{u}^3.
\end{equation} 
 The present-day radiation content of the Universe can be described in terms of the energy density associated with each relativistic particle species at present. This radiation consists of photons and neutrinos plus any additional hidden components, which we call dark radiation (DR):
\begin{eqnarray}
\rho_{radiation}=\rho_{photon}+\rho_{neutrino}+\rho_{DR}\;.
\end{eqnarray}
We can express this in terms of an effective number of neutrino species, $N_{\text{eff}}$, as follows
\begin{equation}
\rho_{radiation}=\rho_{photon}\left(1+\frac{7}{8}\left(\frac{4}{11}\right)^{4/3} N_{\text{eff}}\right)~.
\end{equation}
Any excess can be accounted for by the presence of DR  which is expressed as
\begin{equation}
\rho_{DR}=\rho_{photon}\left(\frac{7}{8}\left(\frac{4}{11}\right)^{4/3} \Delta N_{\text{eff}}\right)~,
\end{equation}
 \begin{figure}[t]
 	\centering \includegraphics[width=9.85cm]{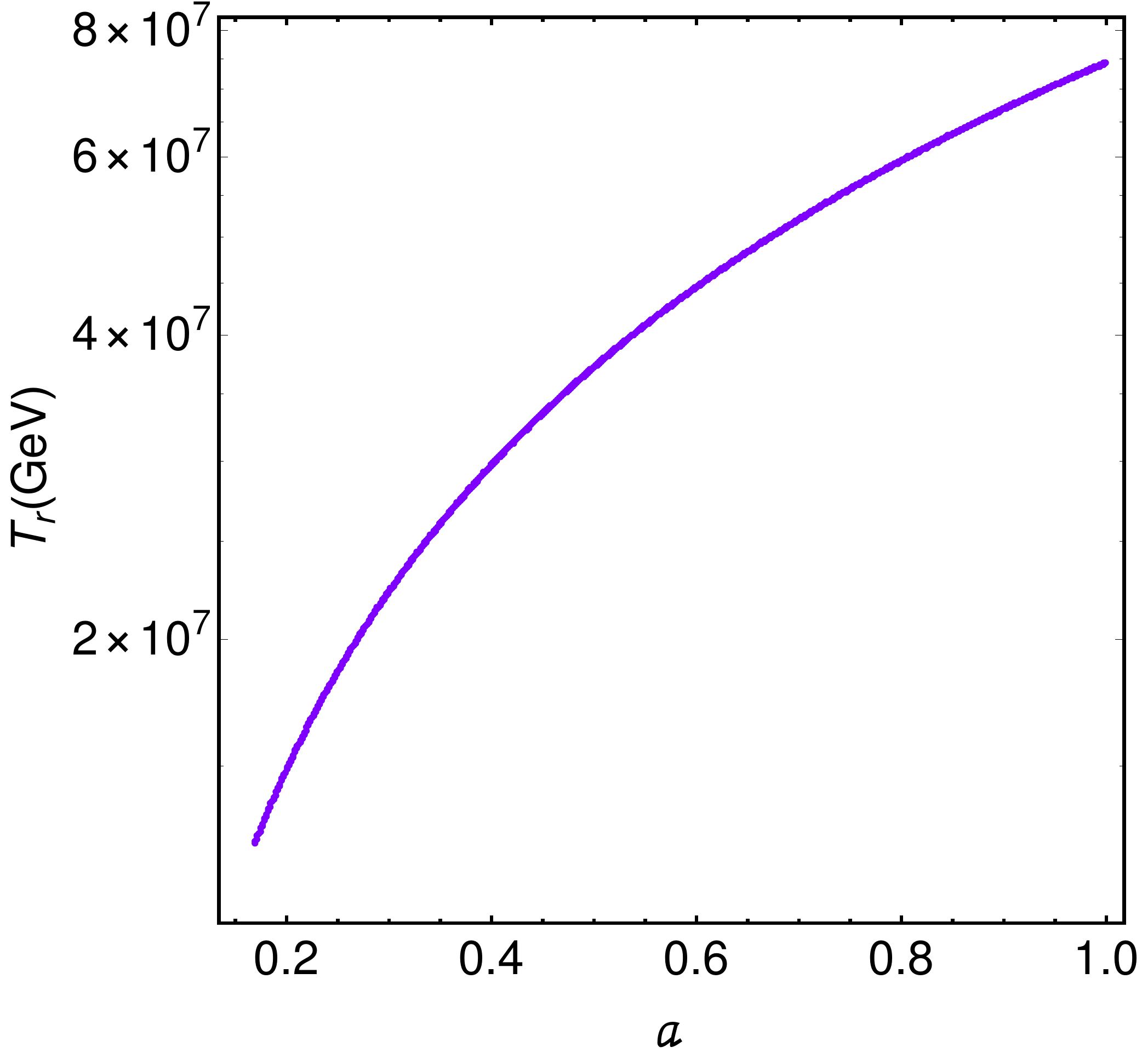}
 	\caption{\small Variations of the reheating temperature ($T_r$) with respect to coefficient $a$ consistent with dark radiation constraint ($\Delta N_{\text{eff}}\lesssim 0.95$) at $95\%$ confidence level.}
 	\label{fig:DR}
 \end{figure} 
where $\Delta N_{\text{eff}}=N_{\text{eff}}-3.046$ is the change in the effective number of neutrino species. If there were no dark radiation we would expect to find $N_{\text{eff}}=3.046$ which is slightly greater than 3 to account for partial reheating due to $e^{+}e^{-}$ annihilation. $\Delta N_{\text{eff}}$ can also be written in terms of decay rate channels
\begin{equation}
\Delta N_{\rm eff} =\frac{43}{7} \left(\frac{10.75}{g_*(T_r)} \right)^{\frac 13}  \frac{\Gamma_{\tau\to DR}}{\Gamma_{\tau\to SM}}=\frac{43}{7} \left(\frac{10.75}{g_*(T_r)} \right)^{\frac 13}  \frac{1}{72 a^2 \lambda^2}~,
\label{dnef}
\end{equation}
where $g_*(T_Rh)$ is the effective degree of freedom at the time of reheating the Universe.
The measured values of $N_{\text{eff}}$ require $\Delta N_{\text{eff}}\lesssim 0.95$ at the $95\%$ confidence level, which translates into a bound on the parameters of the model, $a\cdot\lambda\gtrsim 0.1688$.  In Fig. \ref{fig:results1}, the parametric space predicted by the inflationary analysis has  $0.86\leq a \leq 1$. This range aligns with the values predicted by $\Delta N_{\text{eff}}\lesssim 0.95$ and is consistent with dark radiation. Within this parametric space, we obtain a tensor-to-scalar ratio values of $r\leq 1.4 \times 10^{-4}$,  $1 \times 10^{15} \lesssim M \lesssim 1.3 \times 10^{16}$ GeV, and a soft mass parameter $4.6\times 10^{5} \lesssim M_{s_{c}}\lesssim 1 \times 10^{11}$ GeV.

The isotropy of the cosmic microwave background (CMB) over large scales can be explained by inflation, which is followed by a period of reheating. During this process, the expansion rate slows, leading to the transfer of energy to SM particles that subsequently enter local thermal equilibrium. We consider the scenario where the Universe is reheated by a modulus $u$ decaying into SM particles. In this case the reheating temperature $T_r$, using Eq.(\ref{gammau}) and Eq.(\ref{dnef}) is defined as
\begin{equation}
T_{r}=\sqrt{\Gamma_{u}}=\sqrt{\frac{63}{344 \pi}\Delta N_{\rm eff}\left(\frac{g_*(T_r)}{10.75}\right)^{1/2}a^2 \lambda^2 m_u^3}\;.
\end{equation}
For a SUSY scale at the TeV regime, the constraint on reheating temperature is $T_r\lesssim1$ GeV, which corresponds to $g_*(T_r)= 224/7$, see Ref~\cite{Hebecker:2014gka, Cicoli:2022uqa}. For the present model, we have ${\cal V}_{o}\sim 3.2\times 10^4$ which corresponds to a SUSY scale: $m_{soft}> 10$ TeV. {As a result, the constraint on the reheating temperature is relaxed in this scenario. For SUSY scale greater than 10 TeV we have $g_*(T_r)=106.75$, so the reheating temperature is $T_r\sim 10^7$ GeV as shown in Fig.} \ref{fig:DR}.

\section{Conclusions} \label{sec5}

  We investigated the cosmological and low energy supersymmetry implications of an effective model~\cite{Antoniadis:2018hqy}
stemming from a geometric configuration of intersecting three  D7-branes stacks, within the framework of type-IIB string theory. 
 In this model perturbative string loop corrections which depend logarithmically on the compactification volume ${\cal V}$,  and D-terms
 associated with the universal $U(1)$ factors of D7-brane stacks generate an effective scalar potential with de Sitter vacuum, 
 and stabilize all three K\"ahler moduli fields of the specific geometric setting.  In the present work we took into account the effects of ordinary matter contributions in the K\"ahler potential 
 of the effective model and, in particular, we focused on the role of a generic pair of Higgs fields $\Phi_{1}, \Phi_{2}$ (related to the gauge group of the effective theory) on low energy phenomenology predictions and various cosmological observables. We included matter field content and soft-term contributions as well as Coleman-Weinberg corrections to the previously 
 derived potential and studied the implementation of the standard hybrid inflationary scenario where a singlet gauge field sharing common couplings with the Higgs fields in the superpotential plays the role of the 
 inflaton whilst the Higgs states act as waterfall fields.
 Fixing the spectral index at central value, $n_s=0.96655$, we provided predictions for the 
 remaining cosmological observables in accordance with the latest Planck data.  In particular, we predicted the value of the tensor-to-scalar ratio to be $r\sim{2\times{10^{-4}}}$
 which is much smaller than the current experimental bounds, however, within the reach of future designed experiments.
 Next,  we discussed the decay of the lighter K\"ahler moduli after the end of inflation, which includes modes to visible as well as invisible particles. In particular, in the context of an MSSM effective theory and the presence of a  Giudice-Masiero coupling, the dominant decay of the lightest modulus is to the Higgs fields, in accordance with previous computations~\cite{Hebecker:2014gka}.  
Furthermore, we investigated predictions of the model to dark radiation production and we found  $\Delta N_{\rm eff}\leq{0.95}$ at $2\sigma$ confidence level. This  requires the model parameters $a$ and $\lambda$ (associated with the couplings $\propto a\lambda(\Phi_{1}\Phi_{2}$ + \textit{h.c}) in the K\"ahler potential) to satisfy the bound $a\lambda\gtrsim{0.1688}$, which for $\lambda \approx 1$ translates into a bound for $a$ in the perturbative regime.   Regarding other vital low energy predictions 
  for a volume fixed at  ${\cal V}_o\sim 3.2\times 10^4$ which ensures a dS minimum, {we found that the SUSY scale is greater than 10 TeV. As a result, the constraint imposed on the reheating temperature is relaxed in this scenario. For SUSY scale (>10 TeV) we have $g_*(T_r)=106.75$ ~\cite{Hebecker:2014gka}, and the reheating temperature is $T_r\sim 10^7$ GeV} .
\vfill


\appendix

\section{Soft Term Potential}\label{AppA}
In general, the soft term scalar Lagrangian can be written as
 \begin{equation}
 \mathcal{L}\supset m_\alpha^2 C^{\alpha} \overline{C}^{\bar{\alpha}}
 \end{equation}
where the soft mass $m_\alpha^2$ of the field $C$ defined as
\begin{equation}
m^{2}_\alpha=(m^{2}_{3/2}+ V_{0})-F^{\bar{m}}F^{n}\partial_{\bar{m}}\partial_{n} \log \tilde{K}_{\alpha}.
\end{equation}
Here, $F^{\bar{m}}$ is defined as $F^{\bar{m}}=e^{G/2}K^{\bar{m}n}\partial G/\partial n$.
Utilizing the superpotential and  K\"ahler metric as given in Eq.(2.4) and  Eq.(2.9),  we find that the soft mass of the canonically normalized field s is
 \begin{equation}
 m^{2}_{s}\sim(m^{2}_{3/2}+ V_{0})-\frac{2}{3{\cal V}_{o}^{2}}W_o^2+\cdots
 \end{equation}
where $V_{0}$ represents the minima of the potential. Hence the soft-term potential is 
\begin{equation}\label{B4}
\Delta V_{soft} =  m^{2}_{S}S^{2} = \left[(m^{2}_{3/2}+ V_{0})-\frac{2}{3{\cal V}_{o}^{2}}W_o^2+\cdots\right] S^{2}.
\end{equation}
Taking into account the relation $s^{2}=\frac{(3a\gamma)}{{\cal{V}}_{0}^{2/3}} S^{2}$ between canonically normalized field s and non-canonically normalized field S we have 
\begin{equation}\label{B5}
\Delta V_{soft} = \left[(m^{2}_{3/2}+ V_{0})-\frac{2}{3{\cal V}_{o}^{2}}W_o^2+\cdots\right] \frac{{\cal{V}}_{0}^{2/3}}{(3a\gamma)} s^{2}.
\end{equation}
If we further define the ratio $y=s/M$, then \ref{B4} takes the form 
 \begin{equation}
\Delta V_{\text{soft}}=\left[(m^{2}_{3/2}+ V_{0})-\frac{2}{3{\cal V}_{o}^{2}}W_o^2+\cdots\right]M^2 y^2{\frac{{\cal V}_{o}^{2/3}}{3 a \gamma}}=M_{s_c}^2y^2~,
 \end{equation}
where  $$M_{s_{c}}=\left[(m^{2}_{3/2}+ V_{0})-\frac{2}{3{\cal V}_{o}^{2}}W_o^2+\cdots\right]M^2 {\frac{{\cal V}_{o}^{2/3}}{3 a \gamma}}~.$$

\section{Coleman-Weinberg Corrections}\label{AppB}
{In this Appendix we present analytically the computation of the CW corrections. Using the effective potential
  \begin{equation}
  \small 
V_{\text{eff}}= \frac{\kappa ^2  \alpha \mid\left(M^2-\Phi_{1}\Phi_{2}\right)\mid^2+2\gamma \kappa ^2  S^2 (\alpha\Phi_{1}\Phi_{1}^{\dagger}+\beta \Phi_{2} \Phi_{2}^{\dagger})}{3 a  \alpha \beta \gamma  \mathcal{V}^{4/3}}+\frac{3 W_0^2 (2 \eta_{0} \log (\mathcal{V})-8 \eta_{0}+\xi_{0} )}{2 \mathcal{V}^3}+\frac{d}{\mathcal{V}^2},
  \end{equation}
the scalar mass matrix along the inflationary track read as,
\begin{equation}
M_{S}^2= \begin{pmatrix}
 \frac{\kappa ^2 S^2}{6 a \beta  \mathcal{V}^{4/3}} & 0 & -\frac{\kappa ^2 M^2}{3 a \gamma  \mathcal{V}^{4/3}} & 0 & 0 & 0 & 0 \\
 0 & \frac{\kappa ^2 S^2}{6 a \beta  \mathcal{V}^{4/3}} & 0 & \frac{\kappa ^2 M^2}{3 a \gamma  \mathcal{V}^{4/3}} & 0 & 0 & 0 \\
 -\frac{\kappa ^2 M^2}{3 a \gamma  \mathcal{V}^{4/3}} & 0 & \frac{\kappa ^2 S^2}{6 a \alpha  \mathcal{V}^{4/3}} & 0 & 0 & 0 & 0 \\
 0 & \frac{\kappa ^2 M^2}{3 a \gamma  \mathcal{V}^{4/3}} & 0 & \frac{\kappa ^2 S^2}{6 a \alpha  \mathcal{V}^{4/3}} & 0 & 0 & 0 \\
 0 & 0 & 0 & 0 & 0 & 0 & 0 \\
 0 & 0 & 0 & 0 & 0 & 0 & 0 \\
 0 & 0 & 0 & 0 & 0 & 0 & \frac{28\kappa^2M^4}{27a\gamma\mathcal{V}^{\frac{10}{3}}}+\cdots \\
\end{pmatrix}
\end{equation} 
After diagonalization and choosing $\alpha=\beta$ the scalar mass spectrum  is
\begin{equation}
\small
M_{S}^2=\left[0,0,\frac{\kappa^2 M^2\left(y^2-\frac{2\alpha}{\gamma}\right)}{6a \mathcal{V}^{4/3}\alpha},\frac{\kappa^2 M^2\left(y^2-\frac{2\alpha}{\gamma}\right)}{6a \mathcal{V}^{4/3}\alpha},\frac{\kappa^2 M^2\left(y^2+\frac{2\alpha}{\gamma}\right)}{6a \mathcal{V}^{4/3}\alpha},\frac{\kappa^2 M^2\left(y^2+\frac{2\alpha}{\gamma}\right)}{6a \mathcal{V}^{4/3}\alpha}, \quad \frac{28 \kappa ^2 M^4}{27 a \gamma  \mathcal{V}^{10/3}}+\cdots\right]
\end{equation}
where $y^2=s^2/M^2$. Similarly, the generic fermionic mass matrix is defined as
\begin{equation}
M_{F_{ij}}=e^{{\cal K}/2}\left(W_{ij}+{\cal K}_{ij}W+{\cal K}_{i}W_{j}+{\cal K}_{j}W_{i}+{\cal K}_{i}{\cal K}_{j}W-{\cal K}^{k\bar{l}}{\cal K}_{ij\bar{l}}D_{k}W\right).
\end{equation}
Using the superpotential \eqref{superpotential} and the K\"ahler potential in \eqref{kahlermp1} we obtain
\begin{equation}
M_{F_{ij}}=\begin{pmatrix}
-\frac{6 a \gamma  \kappa M^2 S}{\mathcal{V}^{2/3}}+\frac{9 a^2 \gamma ^2 S^2 }{\mathcal{V}^{4/3}}+\cdots & 0 & 0 & \frac{2 k M^2}{\mathcal{V}} + \frac{2 a \gamma  S \left(k M^2 S-4\right)}{\mathcal{V}^{5/3}}+\cdots \\
 0 & 0 & \kappa S & 0  \\
 0 & \kappa S & 0 & 0  \\
\frac{2 \kappa M^2}{\mathcal{V}} + \frac{2 a \gamma  S \left(\kappa M^2 S-4\right)}{\mathcal{V}^{5/3}}+\cdots  & 0 & 0 & \frac{6}{\mathcal{V}^2}+\frac{34}{3} a \gamma  S^2 \left(\frac{1}{\mathcal{V}}\right)^{8/3}+\cdots \\

\end{pmatrix}.
\end{equation}
After diagonalization, the fermionic masses are
\begin{equation}
 M_{F_{ij}}^2=\left[\kappa^2 M^2y^2, \kappa^2 M^2 y^2, \frac{36a^2\kappa^2 M^6y^2\gamma^2}{\mathcal{V}^{4/3}}+\cdots,\frac{36a^2\kappa^2 M^6y^2\gamma^2}{\mathcal{V}^{4/3}}+\cdots\right].
\end{equation}
The fermionic mass $\frac{36a^2\kappa^2 M^6 y^2\gamma^2}{\mathcal{V}^{4/3}}$ and the bosonic mass $\, \frac{28 \kappa ^2 M^4}{27 a \gamma  \mathcal{V}^{10/3}}$ are small compared to other masses, so their contribution to the Coleman-Weinberg potential is suppressed. Additionally, we focus on the region where $(y^2 \gg  2\alpha/\gamma)$. The relation between canonically and non-canonically normalised field is $s= S \sqrt{3a\gamma}/ V_{0}^{1/3} $.
The effective contribution of the one-loop radiative corrections can be calculated using the Coleman-Weinberg formula
\begin{equation}
\Delta V_{\text{1-loop}}=\frac{1}{64\pi^2}\left[M_{S}^4\log\left(\frac{M_{S}^2}{Q^2} \right)-2M_{F}^4\log\left(\frac{M_{F}^2}{Q^2}\right)-\frac{3}{2}\left(M_{S}^4-2M_{F}^4\right) \right]
\end{equation}
  \begin{equation}
	\Delta V_{\text{1-loop}}=  \frac{\kappa ^4 M^4y^4}{144\pi ^2 a^2  \gamma^2 {\cal V}_o^{4/3} }\left[F(y)-\left(\frac{ 1}{54 a^2 \alpha^2 \gamma^2}-\frac{3}{2}{\cal V}_o^{8/3}\right)\right],
\end{equation}
 where
 \begin{equation}
 F(y)=\frac{1}{81a^2\alpha^2 \gamma^2}\log \left(\frac{\kappa ^2 M^2 y^2}{27 a^2 \alpha\gamma^2  Q^2 {\cal V}_o^{2/3}} \right)-{\cal V}_o^{8/3}\log\left(\frac{\kappa ^2 M^2 y^2 {\cal V}_o^{2/3}}{3 a \gamma Q^2}\right)~.
\end{equation}
}

\newpage 
\color{black}


\begin{thebibliography}{99}

\bibitem{Palti:2019pca}
E.~Palti,
``The Swampland: Introduction and Review,''
Fortsch. Phys. \textbf{67} (2019) no.6, 1900037
[arXiv:1903.06239 [hep-th]].
		
\bibitem{Agmon:2022thq}
N.~B.~Agmon, A.~Bedroya, M.~J.~Kang and C.~Vafa,
``Lectures on the string landscape and the Swampland,''
[arXiv:2212.06187 [hep-th]].
	
\bibitem{VanRiet:2023pnx}
T.~Van Riet and G.~Zoccarato,
``Beginners lectures on flux compactifications and related Swampland topics,''
[arXiv:2305.01722 [hep-th]].
		
\bibitem{Balasubramanian:2005zx}
V.~Balasubramanian, P.~Berglund, J.~P.~Conlon and F.~Quevedo,
``Systematics of moduli stabilization in Calabi-Yau flux compactifications,''
JHEP \textbf{03} (2005), 007
[arXiv:hep-th/0502058 [hep-th]].
			
\bibitem{Berg:2007wt}
M.~Berg, M.~Haack and E.~Pajer,
``Jumping Through Loops: On Soft Terms from Large Volume Compactifications,''
JHEP \textbf{09} (2007), 031
[arXiv:0704.0737 [hep-th]].
		
\bibitem{Reece:2015qbf}
M.~Reece and W.~Xue,
``SUSY\textquoteright{}s Ladder: reframing sequestering at Large Volume,''
JHEP \textbf{04} (2016), 045
[arXiv:1512.04941 [hep-ph]].
	
\bibitem{Cicoli:2017shd}
M.~Cicoli, I.~Garc\`\i{}a-Etxebarria, C.~Mayrhofer, F.~Quevedo, P.~Shukla and R.~Valandro,
``Global Orientifolded Quivers with Inflation,''
JHEP \textbf{11} (2017), 134
[arXiv:1706.06128 [hep-th]].

\bibitem{Cicoli:2021dhg}
M.~Cicoli, I.~G.~Etxebarria, F.~Quevedo, A.~Schachner, P.~Shukla and R.~Valandro,
``The Standard Model quiver in de Sitter string compactifications,''
JHEP \textbf{08} (2021), 109
[arXiv:2106.11964 [hep-th]].

\bibitem{Gao:2022fdi}
X.~Gao, A.~Hebecker, S.~Schreyer and G.~Venken,
``The LVS parametric tadpole constraint,''
JHEP \textbf{07} (2022), 056
[arXiv:2202.04087 [hep-th]].
	
\bibitem{Antoniadis:2018hqy}
I.~Antoniadis, Y.~Chen and G.~K.~Leontaris,
``Perturbative moduli stabilization in type IIB/F-theory framework,''
Eur. Phys. J. C \textbf{78} (2018) no.9, 766
[arXiv:1803.08941 [hep-th]].
	
\bibitem{Antoniadis:2019rkh}
I.~Antoniadis, Y.~Chen and G.~K.~Leontaris,
``Logarithmic loop corrections, moduli stabilization and de Sitter vacua in string theory,''
JHEP \textbf{01}, 149 (2020)
[arXiv:1909.10525 [hep-th]].

\bibitem{Basiouris:2020jgp}
V.~Basiouris and G.~K.~Leontaris,
``Note on de Sitter vacua from perturbative and non-perturbative dynamics in type IIB/F-theory compactifications,''
Phys. Lett. B \textbf{810} (2020), 135809
[arXiv:2007.15423 [hep-th]].

\bibitem{Basiouris:2021sdf}
V.~Basiouris and G.~K.~Leontaris,
``Remarks on the Effects of Quantum Corrections on Moduli Stabilization and de Sitter Vacua in Type IIB String Theory,''
Fortsch. Phys. \textbf{70} (2022) no.2-3, 2100181
[arXiv:2109.08421 [hep-th]].

\bibitem{Cicoli:2023opf}
M.~Cicoli, J.~P.~Conlon, A.~Maharana, S.~Parameswaran, F.~Quevedo and I.~Zavala,
``String cosmology: From the early universe to today,''
Phys. Rept. \textbf{1059} (2024), 1-155
[arXiv:2303.04819 [hep-th]].


\bibitem{Leontaris:2023obe}
G.~K.~Leontaris and P.~Shukla,
``Seeking de Sitter vacua in the string landscape,''
PoS \textbf{CORFU2022} (2023), 058
[arXiv:2303.16689 [hep-th]].



\bibitem{Burgess:2003ic}
C.~P.~Burgess, R.~Kallosh and F.~Quevedo,
``De Sitter string vacua from supersymmetric D terms,''
JHEP \textbf{10}, 056 (2003)
[arXiv:hep-th/0309187 [hep-th]].

 \bibitem{Antoniadis:2021lhi}
 I.~Antoniadis, O.~Lacombe and G.~K.~Leontaris,
 ``Hybrid inflation and waterfall field in string theory from D7-branes,''
 JHEP \textbf{01} (2022), 011
 [arXiv:2109.03243 [hep-th]].
 		 		
\bibitem{Antoniadis:ijmpa}
I.~Antoniadis, O.~Lacombe and G.~K.~Leontaris,
``Type IIB moduli stabilization, inflation, and waterfall fields,''
Int. J. Mod. Phys. A \textbf{37} (2022) no.34, 2244001

\bibitem{hybrid}
A.~D.~Linde,
``Hybrid inflation,''
Phys. Rev. D \textbf{49} (1994), 748-754
[arXiv:astro-ph/9307002].
G.~Lazarides and C.~Panagiotakopoulos,
``Smooth hybrid inflation,''
Phys. Rev. D \textbf{52} (1995), R559-R563
[arXiv:hep-ph/9506325 [hep-ph]].

 

\bibitem{Gukov:1999ya}
S.~Gukov, C.~Vafa and E.~Witten,
``CFT's from Calabi-Yau four folds,''
Nucl. Phys. B \textbf{584} (2000), 69-108
[arXiv:hep-th/9906070 [hep-th]].

\bibitem{Antoniadis:2019doc}
I.~Antoniadis, Y.~Chen and G.~K.~Leontaris,
``Moduli stabilization and inflation in type IIB/F-theory,''
PoS \textbf{CORFU2018} (2019), 068
[arXiv:1901.05075 [hep-th]].
 		
\bibitem{Leontaris:2022rzj}
G.~K.~Leontaris and P.~Shukla,
``Stabilising all K\"ahler moduli in perturbative LVS,''
JHEP \textbf{07} (2022), 047
[arXiv:2203.03362 [hep-th]].


\bibitem{Kreuzer:2000xy}
M.~Kreuzer and H.~Skarke,
``Complete classification of reflexive polyhedra in four-dimensions,''
Adv. Theor. Math. Phys. \textbf{4} (2000), 1209-1230
[arXiv:hep-th/0002240 [hep-th]].

\bibitem{Altman:2014bfa}
R.~Altman, J.~Gray, Y.~H.~He, V.~Jejjala and B.~D.~Nelson,
``A Calabi-Yau Database: Threefolds Constructed from the Kreuzer-Skarke List,''
JHEP \textbf{02} (2015), 158
[arXiv:1411.1418 [hep-th]].


  			 			 	
\bibitem{Blumenhagen:2009gk}
R.~Blumenhagen, J.~P.~Conlon, S.~Krippendorf, S.~Moster and F.~Quevedo,
``SUSY Breaking in Local String/F-Theory Models,''
JHEP \textbf{09} (2009), 007
doi:10.1088/1126-6708/2009/09/007
[arXiv:0906.3297 [hep-th]].
 	
\bibitem{Aparicio:2008wh}
L.~Aparicio, D.~G.~Cerdeno and L.~E.~Ibanez,
``Modulus-dominated SUSY-breaking soft terms in F-theory and their test at LHC,''
JHEP \textbf{07} (2008), 099
doi:10.1088/1126-6708/2008/07/099
[arXiv:0805.2943 [hep-ph]].



\bibitem{Lust:2004fi}
D.~L\"ust, S.~Reffert and S.~Stieberger,
``Flux-induced soft supersymmetry breaking in chiral type IIB orientifolds with D3 / D7-branes,''
Nucl. Phys. B \textbf{706} (2005), 3-52
[arXiv:hep-th/0406092 [hep-th]].


\bibitem{Becker:2002nn}
K.~Becker, M.~Becker, M.~Haack and J.~Louis,
``Supersymmetry breaking and alpha-prime corrections to flux induced potentials,''
JHEP \textbf{06}, 060 (2002)
[arXiv:hep-th/0204254 [hep-th]].



\bibitem{Hebecker:2014gka}
A.~Hebecker, P.~Mangat, F.~Rompineve and L.~T.~Witkowski,
``Dark Radiation predictions from general Large Volume Scenarios,''
JHEP \textbf{09} (2014), 140
doi:10.1007/JHEP09(2014)140
[arXiv:1403.6810 [hep-ph]].

\bibitem{Conlon:2006tj}
J.~P.~Conlon, D.~Cremades and F.~Quevedo,
``Kahler potentials of chiral matter fields for Calabi-Yau string compactifications,''
JHEP \textbf{01} (2007), 022
[arXiv:hep-th/0609180 [hep-th]].

\bibitem{Haack:2006cy}
M.~Haack, D.~Krefl, D.~L\"ust, A.~Van Proeyen and M.~Zagermann,
``Gaugino Condensates and D-terms from D7-branes,''
JHEP \textbf{01} (2007), 078
[arXiv:hep-th/0609211 [hep-th]].

\bibitem{Kachru:2003aw}
S.~Kachru, R.~Kallosh, A.~Linde and S.~P.~Trivedi,
``de Sitter vacua in string theory,''
Phys. Rev. D \textbf{68}, 046005 (2003)
[arXiv:hep-th/0301240 [hep-th]].

\bibitem{Achucarro:2006zf}
A.~Achucarro, B.~de Carlos, J.~A.~Casas and L.~Doplicher,
``De Sitter vacua from uplifting D-terms in effective supergravities from realistic strings,''
JHEP \textbf{06} (2006), 014
[arXiv:hep-th/0601190 [hep-th]].

\bibitem{Antoniadis:2020stf}
I.~Antoniadis, O.~Lacombe and G.~K.~Leontaris,
``Inflation near a metastable de Sitter vacuum from moduli stabilization,''
Eur. Phys. J. C \textbf{80}, no.11, 1014 (2020)
[arXiv:2007.10362 [hep-th]].
	
\bibitem{Brignole:1993dj}
A.~Brignole, L.~E.~Ibanez and C.~Munoz,
``Towards a theory of soft terms for the supersymmetric Standard Model,''
Nucl. Phys. B \textbf{422}, 125-171 (1994)
[erratum: Nucl. Phys. B \textbf{436}, 747-748 (1995)]
[arXiv:hep-ph/9308271 [hep-ph]].
 	
\bibitem{Conlon:2006wz}
J.~P.~Conlon, S.~S.~Abdussalam, F.~Quevedo and K.~Suruliz,
``Soft SUSY Breaking Terms for Chiral Matter in IIB String Compactifications,''
JHEP \textbf{01}, 032 (2007)
[arXiv:hep-th/0610129 [hep-th]].





\bibitem{Demirtas:2020ffz}
M.~Demirtas, M.~Kim, L.~McAllister and J.~Moritz,
``Conifold Vacua with Small Flux Superpotential,''
Fortsch. Phys. \textbf{68} (2020), 2000085
[arXiv:2009.03312 [hep-th]].



\bibitem{Coleman-Weinberg: 1973}
S.R.~Coleman and E.J.~Weinberg,
``Radiative Corrections as the Origin of Spontaneous Symmetry Breaking,''
Phys.\ Rev.\ D\  {\bf 7}, 1888 (1973)
[arXiv:hep-ph/].

\bibitem{Antoniadis:2018ngr}
I.~Antoniadis, Y.~Chen and G.~K.~Leontaris,
Int. J. Mod. Phys. A \textbf{34} (2019) no.08, 1950042
doi:10.1142/S0217751X19500428
[arXiv:1810.05060 [hep-th]].
 	 		 
\bibitem{Planck:2018jri}
Y.~Akrami \textit{et al.} [Planck],
``Planck 2018 results. X. Constraints on inflation,''
Astron. Astrophys. \textbf{641} (2020), A10
[arXiv:1807.06211 [astro-ph.CO]].

\bibitem{Cicoli:2022uqa}
M.~Cicoli, K.~Sinha and R.~Wiley Deal,
``The dark universe after reheating in string inflation,''
JHEP \textbf{12} (2022), 068
[arXiv:2208.01017 [hep-th]].

\bibitem{Cicoli:2011yy}
M.~Cicoli, C.~P.~Burgess and F.~Quevedo,
``Anisotropic Modulus Stabilisation: Strings at LHC Scales with Micron-sized Extra Dimensions,''
JHEP \textbf{10}, 119 (2011)
[arXiv:1105.2107 [hep-th]].

\bibitem{Giudice:1988yz}
G.~F.~Giudice and A.~Masiero,
``A Natural Solution to the mu Problem in Supergravity Theories,''
Phys. Lett. B \textbf{206} (1988), 480-484


\bibitem{Angus:2013zfa}
S.~Angus, J.~P.~Conlon, U.~Haisch and A.~J.~Powell,
``Loop corrections to $\Delta N_{eff}$ in large volume models,''
JHEP \textbf{12} (2013), 061
[arXiv:1305.4128 [hep-ph]].
















\end{thebibliography}
\end{document}